\documentclass[12pt]{article}
\usepackage{amsfonts,amsmath,amsthm,amssymb}
\numberwithin{equation}{section}
\begin{document}

\title{Topological BPS charges in 10 and 11-dimensional supergravity}
\author{Andrew K. Callister\footnote{E-mail: a.k.callister@durham.ac.uk} and Douglas J. Smith\footnote{E-mail: douglas.smith@durham.ac.uk}\\ 
}

\maketitle

\begin{center}

{\em Department of Mathematical Sciences,
University of Durham,
Science Laboratories,
South Road,
Durham. DH1 3LE.
UK}

\end{center}

\vspace{1.4cm}

\begin{abstract}We consider the supersymmetry algebras of the 10 and 11 dimensional maximal supergravities. We construct expressions from which the topological charge structure of the algebras can be determined in supersymmetric curved backgrounds. These are interpreted as the topological charges of the 1/2-BPS states that are found in the theories. We consider charges for all the M-, NS- and D-branes as well as the Kaluza Klein monopoles. We also show that the dimensional reduction relations between the 11-d and IIA charges, and T-duality relations of the IIA and IIB charges match those found for the branes themselves. Finally we consider the massive versions of the IIA and 11-d theories and find that the expressions for the charges, with a slight modification, are still valid in those instances.
\end{abstract}

\section{Introduction}

Over the last decade or so various works have examined the 1/2-BPS states (or branes) of 10 and 11 dimensional supergravity. Since these supergravity theories are the low energy effective field theories of the various superstring theories such states are important because they offer insights into the corresponding superstring theory. Various aspects of the states have been studied including the worldvolume actions, tensions, projection conditions and duality relations. In this paper we consider the branes from the target space perspective and determine the structure of closed expressions that should correspond to the topological charges of the branes in curved backgrounds.\\
\indent It is well known that the flatspace SUSY algebras of supergravity theories receive extensions corresponding to the topological charges of the various p-branes that couple to the theory.  In \cite{deAzcarraga:1989gm} some examples were explicitly calculated. The general result was that for a $p+1$ dimensional extended object, or $p$-brane, the SUSY algebra received an extension of the form
\begin{eqnarray}\label{eq:alg example}
\frac{1}{p!}(C\Gamma_{\mu_1\ldots\mu_p})_{\alpha\beta}Z^{\mu_1\ldots\mu_p} 
\end{eqnarray}
where
\begin{eqnarray}\label{eq:top charge}
Z^{\mu_1\ldots\mu_p}=Q_{(p)}\int dX^{\mu_1}\wedge dX^{\mu_2}\wedge \ldots \wedge dX^{\mu_p}  
\end{eqnarray}
is the topological charge. Here $C$ is the charge conjugation matrix, $X^{\mu}$ are spacetime co-ordinates and $\Gamma$ is an antisymmetric combination of Dirac $\Gamma$ matrices. The integral is understood to be taken over the spatial hypersurface occupied by the brane. $Q_{(p)}$ corresponds to the charge of the brane and is determined by performing an appropriate asymptotic flux integral over a region that surrounds the brane. This is only non-zero if the brane wraps a non-trivial cycle, which gives the charge its topological nature. For the 1/2-BPS states considered in this paper we have the condition that the brane tension $T$ is equal to the fundamental value of $Q_{(p)}$, i.e. the charge of a `single' brane. The above expression can then be re-written with $T$ replacing $Q_{(p)}$ and with the integrand understood to be the pullback to the brane worldvolume.\\
\indent Note that only the spatial components of the charges are associated with the charge of a $p$-brane. For the flatspace case, those components with a time index are interpreted in terms of the Hodge dual of the charge and are therefore the charge components of a $D-p+1$ dimensional extended object where $D$ is the dimension of the spacetime. Such ideas were discussed in \cite{Hull:1997kt,Townsend:1997wg}.\\
\indent In \cite{deAzcarraga:1989gm} it was further shown that the origin of these charges appearing in the SUSY algebra was due to the fact that the Wess-Zumino term in the worldvolume actions is only quasi-invariant under global SUSY transformations. Subsequent work carried out in \cite{Sorokin:1997ps} showed that the presence of a worldvolume gauge potential in the worldvolume action, with non-trivial SUSY transformations, also gives rise to topological terms in the SUSY algebras. The M5-brane example was considered and in this instance a further result showed that the M5-brane algebra not only included a 5-form charge but also the M2-charge. In such a case (\ref{eq:top charge}) with $p=5$ can be thought of as some `core' charge for the M5-brane but another term also appears in the algebra involving the worldvolume gauge potential and the M2-brane charge. We will describe these terms when we consider the M5-brane algebra explicitly.  The general formulation of the method by which a worldvolume gauge potential can give rise to a topological extension to the SUSY algebra was presented in \cite{Hammer:1997ts} where the method was explicitly carried out for the cases of the D-branes. It was found that also in these cases the algebras do not only contain the `core' D-brane charges but also terms with lower rank charges corresponding to the lower dimensional D-branes.\\
\indent The above analyses all assumed a flat background. However, specific cases of curved backgrounds have also been investigated in \cite{Sato:1998yx,Sato:1998ax,Sato:1998yu,Furuuchi:1999tn}. Here various examples of branes immersed in backgrounds sourced by other branes were considered and the worldvolume and spacetime superalgebras were constructed for these cases. See also \cite{Meessen:2003yi,Peeters:2003vz,Lee:2004jx,AliAkbari:2005is} for extensions of AdS spacetime superalgebras. A general method for the construction of the charges in arbitrary supersymmetric curved backgrounds has been given in \cite{Bergshoeff:1998ha} which involved finding a worldvolume action which is invariant under any isometries of the background. The action relevant for the D-branes was explicitly given.\\
\indent In this paper however we follow the example of \cite{Hackett-Jones:2003vz}. Here general expressions were constructed from which the charges in any specific supersymmetric, curved background could be calculated. We will refer to these expressions as the generalised charges or as simply just the charges. Essentially the method involved finding expressions that were closed (and hence topological) for general background field configurations and that also simplify in flatspace so that the standard flatspace algebra can be read off. As well as involving the background field potentials these expressions also contain bilinear forms made out of products of a Killing spinor and antisymmetric combinations of Dirac $\Gamma$ matrices. Some degree of supersymmetry must therefore be preserved since it is assumed a Killing spinor exists. Such bilinears appear when one calculates the commutator of a specific SUSY transformation and it is in this way that these closed expressions are related to the charges in the algebra. Since the closed expressions apply to arbitrary supersymmetric backgrounds, they can be used to determine the charges in the SUSY algebras in such backgrounds. It is these closed expressions that we refer to as the generalised charges. We do not consider in any detail the role of the worldvolume fields in these charges since we are considering the algebra from the target space perspective. Furthermore, we only consider bosonic charges by assuming all the fermionic fields are set to zero.\\
\indent In \cite{Hackett-Jones:2003vz} cases of the M2 and M5-branes in 11 dimensional supergravity were considered, and the generalised charges were determined explicitly. Subsequently partial analyses have been carried out for the IIA \cite{Saffin:2004ar} and IIB \cite{Hackett-Jones:2004yi} cases. In this paper, we consider the IIA, IIB and 11 dimensional theories and construct the generalised charges for the `standard' spectrum of states that have charges in the flat space algebra. This includes the M-, NS- and D-branes as well as the Kaluza-Klein (KK) monopoles which are purely gravitational states. To properly investigate the D8-brane and M9-brane cases we must consider the massive versions of IIA and 11-d supergravity. We also examine the T-duality relations between the IIA and IIB charges, as well as mentioning the reduction relations between the 11-d and IIA charges.\\
\indent The organisation of this paper is as follows: in Section \ref{sec:11 dimensional supergravity} we review the work of \cite{Hackett-Jones:2003vz} explaining the steps to go about determining the generalised charges and explicitly giving the M2 and M5-brane charges. Then in Sections \ref{sec:massless IIA supergravity} and \ref{sec:type IIB supergravity} we apply the method to the massless IIA and IIB branes respectively for the D- and NS-branes. We then explain how to find the generalised charges for the KK monopoles in Section \ref{sec:Kaluza-Klein monopole charges}. We then consider the massive versions of the IIA and 11 dimensional theories in Section \ref{sec:massive supergravity} and find how the massless generalised charges should be modified for those cases, and then go on to present the M9-brane charge. In Section \ref{sec:T-dual} we write down the T-duality rules for all the fields and show the T-duality relations between the charges. In Appendix \ref{sec:conventions} we state our conventions while in Appendix \ref{sec:reducs} we give the reduction rules from 11 dimensions to IIA. Finally in Appendix \ref{sec:summary} we collect all our expressions for the generalised charges for easy reference.

\section{11 dimensional Supergravity}\label{sec:11 dimensional supergravity}
We begin our formulation of the generalised charges by considering the standard branes found in 11-d supergravity, namely the M2 and M5-branes. After giving a brief description of these branes to illustrate the essential characteristics, we will consider the SUSY algebra and demonstrate how one can construct the generalised charges. This work is essentially a review of \cite{Hackett-Jones:2003vz}. We will also compare the charges with previous results.  

\subsection{Branes in 11-dimensional Supergravity}
In our conventions, the bosonic part of the 11-dimensional supergravity action is given by:\footnote{We denote 11-dimensional objects with a hat.}
\begin{eqnarray}\label{eq:11-d action}
\hat{S}=\frac{1}{2}\int d^{11}x\sqrt{-\hat{g}} \biggl(\hat{R}-\frac{1}{2.4!}|\hat{F}|^2\biggr)
\end{eqnarray}
together with a Chern-Simons term
\begin{eqnarray}\label{eq:11-d CS}
-\frac{1}{2}\int \frac{1}{6}\hat{F}\wedge \hat{F}\wedge \hat{A}
\end{eqnarray}
where $\hat{F}$ is the 4-from field strength, related to the 3-form potential $\hat{A}$ by $\hat{F}=d\hat{A}$. Varying this action with respect to
$\hat{A}$ gives the sourceless equation of motion:
\begin{eqnarray}
d\hat{F}^{(7)}+\frac{1}{2}\hat{F}\wedge \hat{F}=0
\end{eqnarray}
where $\hat{F}^{(7)}=\ast \hat{F}$. This allows for the definition of a 6-form dual potential, $\hat{C}$, given by
\begin{eqnarray}
d\hat{C}=\hat{F}^{(7)} +\frac{1}{2}\hat{A}\wedge \hat{F}
\end{eqnarray}
\indent The standard brane solutions of this action are the
M2-brane, or supermembrane, given explicitly by:
\begin{eqnarray}\label{eq:M2-brane}
\nonumber ds^2_{(11)}&=&H^{-\frac{2}{3}}dx^2_{(1,2)}+H^{\frac{1}{3}}dx^2_{(8)}\\
\nonumber \hat{F}&=&\pm d(H^{-1})\wedge \epsilon _{(1,2)}\\
H&=&1+\frac{c_{(2)}N}{r^6}
\end{eqnarray}
and the M5-brane given by:
\begin{eqnarray}\label{eq:M5-brane}
\nonumber ds^2_{(11)}&=&H^{-\frac{1}{3}}dx^2_{(1,5)}+H^{\frac{2}{3}}dx^2_{(5)}\\
\nonumber \hat{F}&=&\pm \ast d(H)\\
H&=&1+\frac{c_{(5)}N}{r^3}
\end{eqnarray}
Here $r$ is the radial coordinate on the transverse Euclidean space
and $\ast$ here is the Hodge dual on this transverse space only. The constants $c_{(p)}$ are related to the branes' tensions. These
solutions are each interpreted as $N$ infinite, flat and coincident branes located at
$r=0$. The entire solution depends only on the form of the harmonic
function $H$. This function can in fact be multi-centred and the
solution becomes that of several non-coincident parallel branes.\\
\indent Both brane solutions are charged objects under $F$, the
M2-brane being electrically charged and the M5-brane magnetically
charged.  The charge (number of branes) can be calculated via Gauss' Law by
evaluating the total flux at the asymptotic spatial infinity of the
branes. For the M2-brane this boundary is $S^7$ and so the charge is
given explicitly by
\begin{eqnarray}
Q_{(2)}=\frac{1}{\Omega_7}\int_{S^7}(\hat{F}^{(7)} +\frac{1}{2}\hat{A}\wedge \hat{F})
\end{eqnarray}
where $\Omega_7$ is the volume of
the unit 7-sphere. Similarly for the M5-brane case the boundary is
$S^4$ and the charge is given by
\begin{eqnarray}
Q_{(5)}=\frac{1}{\Omega_4}\int_{S^4} \hat{F}
\end{eqnarray}

\subsection{Charges in 11-dimensional Supergravity}\label{sec:11-d charges}
\indent The flatspace SUSY algebra receives modifications to the right hand side
of a 2-form and 5-form corresponding to the topological charges of the M2 and M5-branes respectively:
\begin{eqnarray}\label{eq:11-d algebra}
\{\hat{Q}_\alpha,\hat{Q}_\beta\}&=&(C\hat{\Gamma} ^{\mu})_{\alpha
\beta}\hat{P}_{\mu}+\frac{1}{2}(C\hat{\Gamma} _{\mu_1\mu_2})_{\alpha
\beta}\hat{Z}^{\mu_1\mu_2}\\ \nonumber &&+\frac{1}{5!}(C\hat{\Gamma} _{\mu_1\ldots \mu_5})_{\alpha \beta}\hat{Z}^{\mu_1\ldots \mu_5}
\end{eqnarray}
where the charges are given by (\ref{eq:top charge}) and $\hat{P}_{\mu}$ is the momentum. As already mentioned, it is only the purely spatial components which are associated with the M2 and M5-brane charges, whereas those components that include a time index are interpreted by taking the Hodge dual of the charges from which one infers the existence of extended objects with 7 and 10 dimensional worldvolumes, as discussed in \cite{Hull:1997kt}. These solutions have been investigated and are known as the Kaluza-Klein (KK) monopole \cite{Sorkin:1983ns,Gross:1983hb,Bergshoeff:1997gy} and the M9-brane \cite{Bergshoeff:1998bs} respectively.  The KK monopole solution consists of a 6 dimensional worldvolume and a 4 dimensional Taub-NUT transverse space. Therefore there is an extra isometry transverse to the worldvolume not found in the standard cases of branes. The M9-brane only exists in the so-called `massive' version of 11-d SUGRA presented in \cite{Bergshoeff:1997ak} which upon dimensional reduction reduces to Romans' massive IIA theory \cite{Romans:1985tz}, and which has an isometry necessarily built into it. New potentials derived from the Killing vector associated with the isometry and from the mass parameter are introduced and couple to these `branes'.\\
\indent Constructing the generalised charges for these brane-like objects is made slightly more complicated due to the requirement that the background has an isometry. We will return to these cases in later sections but for now only consider the M2 and M5-brane charges. It is worth noting that the notion of reinterpreting the time components of the charges appearing in (\ref{eq:11-d algebra}) in terms of the spatial components of the Hodge duals of the charges only makes sense in flat backgrounds. For curved backgrounds time and space components are mixed when the charges are Hodge dualised. Therefore generally speaking the algebra (\ref{eq:11-d algebra}) should be written so as to include all the charges (i.e also the KK monopole and M9-brane charges) with the understanding that they are only evaluated on space-like hypersurfaces. This is a more democratic treatment of the states.\\
\indent Obviously one cannot write a general algebra that applies to all spacetimes, since the algebra depends on the specific isometries present in a specific spacetime. However generalised expressions can be constructed from which one can deduce the structure of the charges appearing in the algebra in arbitrary supersymmetric backgrounds. These are precisely the generalised charges constructed in \cite{Hackett-Jones:2003vz} for the M2 and M5-branes. We review the construction of these charges here.\\
\indent Firstly, one recalls that it is possible to construct bilinear forms of various ranks out of products of gamma matrices and spinors as follows:
\begin{eqnarray}
K_{(p)}=\overline{\epsilon}\Gamma _{(p)}\epsilon
\end{eqnarray}
where $\Gamma_{(p)}$ is an antisymmetric product of $p$ $\Gamma$
matrices, and $\epsilon$ is a Majorana spinor for the D=11 case here. In principle these forms can be constructed from 2 different Killing spinors, however in this paper we only consider the case where a single Killing spinor is used. In this case the forms are only non zero for certain values of $p$ in a given dimension, depending on whether the product of $\Gamma$ matrices is symmetric or antisymmetric in their spinor indices. In 11 dimensions one obtains non-zero forms for $p= 1,2$ and $5$, which we label as follows:\footnote{Note that bilinears such as these are not fully independent, but rather satisfy certain Fierz identities.}
\begin{eqnarray}\label{eq:11-d bilinears}
\hat{K}_{\mu}&=&\overline{\hat{\epsilon}}\hat{\Gamma}_{\mu} \hat{\epsilon}\\
\hat{\omega}_{\mu_1\mu_2} &=&\overline{\hat{\epsilon}}\hat{\Gamma}_{\mu_1\mu_2} \hat{\epsilon}\\
\hat{\Sigma}_{\mu_1\ldots \mu_5}&=&\overline{\hat{\epsilon}}\hat{\Gamma}_{\mu_1\ldots \mu_5} \hat{\epsilon}
\end{eqnarray}
related to these by Hodge duality there also exist 6, 9 and 10-forms:
\begin{eqnarray}
\hat{\Lambda}_{(6)} &=&\overline{\hat{\epsilon}}\hat{\Gamma}_{(6)} \hat{\epsilon}=\hat{\ast} \hat{\Sigma}_{(5)}\\
\hat{\Pi}_{(9)} &=&\overline{\hat{\epsilon}}\hat{\Gamma}_{(9)} \hat{\epsilon}=-\hat{\ast} \hat{\omega}_{(2)}\\
\hat{\Upsilon}_{(10)}&=&\overline{\hat{\epsilon}}\hat{\Gamma}_{(10)} \hat{\epsilon}=\hat{\ast} \hat{K}_{(1)}
\end{eqnarray}
where the ranks of the forms have been temporarily indicated for convenience. Note that the relative minus signs arise from dualising products of $\Gamma$ matrices, our conventions are outlined in Appendix \ref{sec:conventions}. We consider these higher rank dual bilinears in their own right as we follow a `democratic' view of the theory where any object is considered on equal footing to its dual. It is necessary to do this if we want to construct generalised charges for the KK monopole and M9-brane.\\
\indent These bilinear expressions act as calibrations for the various 1/2-BPS states in flatspace. This can be shown from considering the SUSY algebra and is discussed in \cite{Gutowski:1999tu}. Therefore there is a natural association between the bilinears, 1/2-BPS states and charges that appear in the flatspace SUSY algebra. The situation is summarised in Table \ref{table:11-d states}. We have neglected to include the one-form bilinear $\hat{K}$, associated with M-wave solutions, since the corresponding `charge' in this instance is the momentum $P$. This case is therefore qualitatively different to the other 1/2-BPS states and we do not consider it in this paper.\footnote{In the case of the momentum $P_M$, the time component is the Hamiltonian so the dual `charge' is not considered either.} As we will see however, $\hat{K}$ is Killing so its correspondence with the momentum seems natural, and it plays a key role in the generalised charges.

\begin{table}[ht]
\centering
\begin{tabular}{|c||c|c|c|}
\hline 
BPS state & Charge & Bilinear & Potential\\
\hline \hline
M2 & $\hat{Z}_{i_1i_2}$ & $\hat{\omega}$ & $\hat{A}$\\
\hline
M5 & $\hat{Z}_{i_1\ldots i_5}$ & $\hat{\Sigma}$ & $\hat{C}$\\
\hline
KK monopole & $\hat{\ast}(\hat{Z}_{0i_1\ldots i_4})$ & $\hat{\Lambda}$ & $\hat{N}^{(8)}$\\
\hline
M9 & $\hat{\ast}(\hat{Z}_{0i_1})$ & $\hat{\Pi}$ & $\hat{A}^{(10)}$\\
\hline
\end{tabular}
\caption{Branes, charges and their associated bilinears in the 11-d theory. Also included are the potentials that minimally couple to the branes.}
\label{table:11-d states}
\end{table}

\subsubsection{M2-brane charge}

\indent To see how the bilinears can be linked to the charges in the SUSY algebra we can consider the case of an M2-brane probe in flat space and introduce constant commuting Majorana spinor fields $\hat{\epsilon}^{\alpha}$ to parametrise the SUSY transformations. In this instance, the M2-brane truncation of (\ref{eq:11-d algebra}) leads to
\begin{eqnarray}
\nonumber
\{\hat{\epsilon}^{\alpha}\hat{Q}_{\alpha},\hat{\epsilon}^{\beta}\hat{Q}_{\beta}\}=2(\hat{\epsilon}Q)^2&=&(\hat{\epsilon}^TC\hat{\Gamma}^{\mu}\hat{\epsilon})\hat{P}_{\mu}+\frac{1}{2}(\hat{\epsilon}^TC\hat{\Gamma}_{\mu_1\mu_2}
\hat{\epsilon})\hat{Z}^{\mu_1\mu_2}\\
&=&\hat{K}^{\mu}\hat{P}_{\mu}+\frac{1}{2}\hat{\omega}_{\mu_1\mu_2}\hat{Z}^{\mu_1\mu_2}
\end{eqnarray}
where we have used $C=\hat{\Gamma}_0$ and the definitions for $\hat{K}$ and $\hat{\omega}$ in the second line. The pullback of this to the M2-brane worldvolume gives the M2-brane worldvolume SUSY algebra as
\begin{eqnarray}
2(\hat{\epsilon} \hat{Q})^2=\hat{K}^{\mu}\int_{M2} d^2\sigma
\hat{p}_{\mu}+\frac{1}{2}T\ \hat{\omega}_{\mu_1\mu_2}\int_{M2} d\hat{X}^{\mu_1}\wedge d\hat{X}^{\mu_2}
\end{eqnarray}
thus
\begin{eqnarray}\label{eq:11-d algebra(omega)}
2(\hat{\epsilon} \hat{Q})^2=\int_{M2} d^2\sigma \hat{K}^{\mu}\hat{p}_{\mu}+T\int_{M2}\hat{\omega}
\end{eqnarray}
where we have used the definition of $\hat{Z}^{\mu_1\mu_2}$, (\ref{eq:top charge}), and the fact that the momentum $\hat{P}_{\mu}$ is defined by the integration of the momentum density $\hat{p}_{\mu}(\sigma)$ over the spatial worldvolume of the brane. Note that we have replaced $\hat{Q}_{(2)}$ by the brane tension $T$ since we have performed a pullback to the brane. We have also moved the the constant bilinears into the integrals so that the expression is in the same form as we will find for curved backgrounds. This suggests that $\hat{\omega}$ should be thought of as a charge density for the M2-brane. $\hat{\omega}$ in (\ref{eq:11-d algebra(omega)}) represents a central extension to the SUSY algebra and as a result must be topological. We therefore require $\hat{\omega}$ to be closed. We shall see shortly that in a flat background i.e. when the background fields vanish, this is the case. However, in the more general instance where the the background fields are non-zero $\hat{\omega}$ is found to no longer be closed so it alone cannot account for the full extension to the SUSY algebra. An extra term must be added to the RHS of (\ref{eq:11-d algebra(omega)}) to form an expression that is closed generally. This expression will be our generalised charge (density) for the M2-brane.\\
\indent In order to find this extra term we must first know what $\hat{\omega}$ itself actually differentiates to as a general function of the background field strengths. We restrict our attention to purely bosonic SUSY solutions of the theory. In this case the supersymmetry transformations of the Rarita-Schwinger fermion $\hat{\psi}_{\mu}$ must be zero in order for the solution to remain purely bosonic. Therefore $\hat{\epsilon}$ is no longer constant, but rather it must satisfy the Killing spinor equation: 
\begin{eqnarray}
\delta_{\hat{\epsilon}} \hat{\psi}_{\mu}=\hat{\tilde{D}}_{\mu}\hat{\epsilon}=0
\end{eqnarray}
where
\begin{eqnarray}\label{eq:11-d killingspinor}
\hat{\tilde{D}}_{\mu}=\hat{\nabla}_{\mu}+\frac{1}{288}\biggl[\hat{\Gamma}_{\mu}^{\phantom{\mu}\nu_1\ldots \nu_4}-8
\delta_{\mu}^{\nu_1}\hat{\Gamma}^{\nu_2\nu_3\nu_4}\biggr]\hat{F}_{\nu_1\ldots \nu_4}
\end{eqnarray}
We can construct the bilinear forms from Killing spinors and use (\ref{eq:11-d killingspinor}) to calculate their covariant derivatives according to
\begin{eqnarray}
\nonumber \hat{\nabla}_{\mu}\hat{K}_{(p)\nu_1\ldots \nu_p}&=&\hat{\nabla}_{\mu}(\hat{\overline{\epsilon}}\hat{\Gamma}_{\nu_1\ldots \nu_p}\hat{\epsilon})\\
&=&(\hat{\overline{\nabla_{\mu}\epsilon}})\hat{\Gamma}_{\nu_1\ldots \nu_p}\hat{\epsilon}+\hat{\overline{\epsilon}}
\hat{\Gamma}_{\nu_1\ldots \nu_p}(\hat{\nabla}_{\mu}
\hat{\epsilon})
\end{eqnarray}
where we have used the fact that the $\Gamma$ matrices are covariantly constant. Antisymmetrising then gives differential relations for each of the bilinears. For the M2-brane we only need the relation for $\hat{\omega}$ but we state all the relations now for convenience:
\begin{eqnarray}\label{eq:11-d dK}
d\hat{K}&=&\frac{2}{3}i_{\hat{\omega}}\hat{F}+\frac{1}{3}i_{\hat{\Sigma}}\hat{F}^{(7)}\\
\label{eq:11-d domega} d\hat{\omega} &=&i_{\hat{K}}\hat{F}\\ \label{eq:11-d dSigma}
d\hat{\Sigma}&=&i_{\hat{K}}\hat{F}^{(7)}-\hat{\omega}\wedge \hat{F}\\\label{eq:11-d dLambda}
d{\hat{\Lambda}}_{\mu_1\ldots \mu_7}&=&\frac{14}{3}{\hat{\omega}}^{\nu}_{\phantom{\nu}[\mu_1}\hat{F}^{(7)}_
{\mu_2\ldots \mu_7]\nu}-\frac{35}{3}{\hat{\Sigma}}^{\nu}_{\phantom{\nu}[\mu_1\ldots \mu_4}{\hat{F}}_{\mu_5\mu_6\mu_7]\nu}
\\\label{eq:11-d dPi}
d{\hat{\Pi}}&=&-\frac{1}{3}\hat{F}\wedge \hat{\Lambda}\\\label{eq:11-d dUpsilon}
d{\hat{\Upsilon}}&=&0
\end{eqnarray}
where the first three relations have been derived previously in \cite{Gauntlett:2002fz,Gauntlett:2003wb}. These references also show that $\hat{K}$ is a Killing vector, $\hat{\nabla}_{(\mu}\hat{K}_{\nu)}=0$, and discuss that it must be time-like or null. Furthermore, it is straight forward to show from the Bianchi identity and (\ref{eq:11-d domega}) that 
\begin{eqnarray}\label{eq:11-d F gauge}
{\cal L}_{\hat{K}}\hat{F}=0
\end{eqnarray}
 and so $\hat{K}$ generates a symmetry of the solution.\\
\indent By considering equation (\ref{eq:11-d domega}) one sees that
$\hat{\omega}$ is indeed closed when $\hat{F}=0$, but is not closed in
general. However a 2-form that is generally closed can be
constructed. In order to do this one must make use of the following
identity concerning the Lie derivative of a $p$-form $\alpha$ with
respect to a given Killing vector field $X$:
\begin{eqnarray}\label{eq:lie ident}
{\cal L}_X\alpha=d(i_X\alpha)+i_Xd\alpha
\end{eqnarray}
Then one finds that the expression 
\begin{eqnarray}\label{eq:11-d M2-charge}
\hat{L}_{(2)}=\hat{\omega}+i_{\hat{K}}\hat{A}
\end{eqnarray}
is closed if one chooses a gauge where
\begin{eqnarray}\label{eq:11-d 3-pot gauge}
{\cal L}_{\hat{K}}\hat{A}=0
\end{eqnarray}
We now must consider whether such a gauge condition is possible. From (\ref{eq:11-d F gauge}) we see that ${\cal L}_{\hat{K}}\hat{A}$ is closed and therefore, at least locally, exact. In fact we can show this directly by realising that for a general gauge choice for $\hat{A}$ we have $d\hat{L}_{(2)}={\cal L}_{\hat{K}}\hat{A}$. Then considering the gauge transformation $\hat{A} \rightarrow \hat{A}+d\hat{\lambda}_{(2)}$ we find ${\cal L}_{\hat{K}}\hat{A}\rightarrow {\cal L}_{\hat{K}}\hat{A}+d({\cal L}_{\hat{K}}\hat{\lambda}_{(2)})$, in other words ${\cal L}_{\hat{K}}\hat{A}$ is shifted by an exact amount, which can be shown to be arbitrary by considering the degrees of freedom of $\hat{\lambda}_{(2)}$. Therefore it must be possible in this instance to always satisfy the condition (\ref{eq:11-d 3-pot gauge}).\\
\indent Therefore $\hat{L}_{(2)}$ is the generalised charge for the M2-brane. Obviously the gauge condition (\ref{eq:11-d 3-pot gauge}) is essential to the definition of $\hat{L}_{(2)}$ since if it is not satisfied then $\hat{L}_{(2)}$ is not closed. A charge still exists if (\ref{eq:11-d 3-pot gauge}) is not satisfied, although it takes the more complicated form 
\begin{eqnarray}\label{eq:11-d 2-charge}
\hat{L}_{(2)}=\hat{\omega}+i_{\hat{K}}\hat{A}-i_{\hat{K}}d\hat{\lambda}_{(2)}
\end{eqnarray}
for some $\hat{\lambda}_{(2)}$ that satisfies ${\cal L}_{\hat{K}}d\hat{\lambda}_{(2)}={\cal L}_{\hat{K}}\hat{A}$. Since the functional form of the charge is dependent on the gauge choice, both the charge and gauge choice must both be specified. In the current example we will assume the gauge condition (\ref{eq:11-d 3-pot gauge}) has been fixed, and for the other charges in this paper we will make analogous gauge choices.\\
\indent The commutator of the SUSY transformations for curved spaces that are asymptotically Minkowski should generalise from (\ref{eq:11-d algebra(omega)}) to 
\begin{eqnarray}\label{eq:M2 comm}
2(\hat{\epsilon} \hat{Q})^2=\int_{M2} d^2\sigma \hat{K}^M\hat{p}_M+T\int_{M2}(\hat{\omega}+i_{\hat{K}}\hat{A})
\end{eqnarray}
So unlike for the flat space case, the $\hat{\Gamma}$ matrices and spinors here are not constant and must be brought inside the integral.\\
\indent The form of $\hat{L}_{(2)}$ is essentially the same as the generalised calibration discussed in \cite{Gutowski:1999tu} and this gives a second interpretation of $\hat{L}_{(2)}$. This means that the pullback of $\hat{\omega}$ to any 2-plane is bounded from above by a term proportional to the volume form of that 2-plane, as in the case of standard calibrations. Here though since $\hat{\omega}$ is no longer closed, the surfaces that saturate this bound have minimum energy rather than minimum volume, with the field $\hat{A}$ giving a contribution to the energy.\\ 
\indent We now demonstrate how to deduce worldvolume superalgebras from (\ref{eq:M2 comm}) by reproducing the result found in \cite{Sato:1998yx} where the M2-brane worldvolume algebra was determined in an M2-brane sourced background. To do this we recall that for M2-brane backgrounds the Killing spinors take the form $\hat{\epsilon}=H^{-\frac{1}{6}}\hat{\epsilon}_0$ where $\hat{\epsilon}_0$ is a constant spinor and $H$ is the harmonic function appearing in the M2-brane spacetime, (\ref{eq:M2-brane}). We then simply strip off the constant spinors $\hat{\epsilon}_0$ and absorb the factors of $H$ on the LHS into the SUSY generators which converts them from target space to worldvolume SUSY generators. Then we convert $\hat{\Gamma}_0$, used in the definition of the curved space $\hat{\overline\epsilon}$'s present in the bilinears, to the charge conjugation matrix $\hat{\Gamma}_{\underline{0}}$. The result will precisely coincide with the algebra given in \cite{Sato:1998yx} if we define our potential so that its non-zero components are proportional to $H^{-1}-1$, so that they vanish asymptotically.

\subsubsection{M5-brane charge}
\indent Essentially the same procedure can also be carried out for the M5-brane charge. Here, for a flat spacetime with an M5-brane probe the SUSY algebra (\ref{eq:11-d algebra}) leads to
\begin{eqnarray}\label{eq:11-d algebra(sigma)}
2(\hat{\epsilon} \hat{Q})^2=\int d^2\sigma \hat{K}^M\hat{p}_M+Q_{(5)}\int\hat{\Sigma}
\end{eqnarray}
where it can be seen from (\ref{eq:11-d dSigma}) that $\hat{\Sigma}$ is
also only closed in the absence of any background fields, but not in general. So again, we need to replace this by a 5-form expression that is closed generally.\\
\indent Following the example of generalised calibrations presented in \cite{Gutowski:1999tu} which applied for the M2-brane case, we would expect a term of the form $i_{\hat{K}}\hat{C}$ to appear in the M5-brane generalised charge. This is natural since it is the dual potential $\hat{C}$ to which the M5-brane couples. Adding such a term to $\hat{\Sigma}$ on its own does not however lead to a closed expression and so other terms must be added also.\\
\indent As mentioned in the introduction, in \cite{Sorokin:1997ps} it was shown that even with a flat target space the M5-brane algebra included not only a term of the form of (\ref{eq:top charge}) with $p=5$, but also a term of the form 
\begin{eqnarray}
\frac{1}{2}(\hat{C}\hat{\Gamma}_{\mu_1\mu_2})_{\alpha\beta}\hat{\tilde{Z}}^{\mu_1\mu_2}
\end{eqnarray}
where in this case
\begin{eqnarray}
\hat{\tilde{Z}}^{\mu_1\mu_2}=\int_{M5}d\hat{X}^{\mu_1}\wedge d\hat{X}^{\mu_2}\wedge dB
\end{eqnarray}
Here $B$ is the 2-form worldvolume gauge potential of the M5-brane and $dB=H+A$ where $H$ is the modified worldvolume field strength and $A$ is the pullback of $\hat{A}$ to the brane worldvolume. This term arises due to the possibility of having M2-branes contained entirely within the M5-brane worldvolume, and will be non-zero in such instances. We do not consider the worldvolume field $B$ in this paper,\footnote{Its appearance in the generalised M5-brane charge is proposed in \cite{Hackett-Jones:2003vz} however.} however the presence of the pullback of $\hat{A}$ suggests that the generalised charge should contain a term of the form $\hat{\omega}\wedge \hat{A}$. Indeed we find that this is the case.\\
\indent After some consideration we find that the expression 
\begin{eqnarray} \label{eq:11-d 5-charge}
\hat{L}_{(5)}=\hat{\Sigma}+i_{\hat{K}}\hat{C}+\hat{L}_{(2)}\wedge \hat{A}-\frac{1}{2}\hat{A}\wedge i_{\hat{K}}\hat{A}
\end{eqnarray}
is closed if we once again fix the gauge according to (\ref{eq:11-d 3-pot gauge}) as well as 
\begin{eqnarray}\label{eq:11-d 6-pot gauge}
{\cal L}_{\hat{K}}\hat{C}=0
\end{eqnarray}
The argument used to show that (\ref{eq:11-d 3-pot gauge}) was a possible gauge choice can be used here for the above gauge condition also. In fact similar situations are generally found with the charges. In the rest of this paper we will simply state the relevant gauge choices without repeating this argument.\\
\indent The result (\ref{eq:11-d 5-charge}) shows that the generalised M5-brane charge takes a slightly more complicated form than the M2-brane charge. This is due to the presence of the worldvolume field strength and the possibility of embedding M2-branes within the M5-brane which results in the extra terms present in $\hat{L}_{(5)}$ that have no analogues in the $\hat{L}_{(2)}$ case. Following from the M2-brane example, $\hat{L}_{(5)}$ can act as a generalised calibrating form for 5-planes. Then the structure of $\hat{L}_{(5)}$ shows how each field contributes to the `flux' energy of the brane.\\ 
\indent  Once again it would be fairly straightforward to produce the worldvolume algebras in curved backgrounds, albeit with the worldvolume field $B$ set to zero. Some examples were found in \cite{Sato:1998yx}, for M2 and M5-brane backgrounds. We see the same general structure as the charge presented here, except for those backgrounds the term $\hat{A}\wedge i_{\hat{K}}\hat{A}$ vanishes so is not present in the algebra.

\subsubsection{Other 11-d SUGRA charges}
A natural next step would be to consider the KK monopole and M9-brane states and determine their charges. Naively their charges would simply be the 6-form bilinear $\hat{\Lambda}$ and the 9-form $\hat{\Pi}$ respectively. The full charges for these states must therefore necessarily also include these bilinears. For the case of the KK monopole, finding the charge would essentially amount to finding a collection of terms involving potentials and bilinears that upon differentiation would cancel the terms on the RHS of (\ref{eq:11-d dLambda}). However, examining (\ref{eq:11-d dLambda}) we see that index contractions prevent us from doing this. This problem also prevents us from finding a charge for the M9-brane since it too would be required to contain a term involving $\hat{\Lambda}$. Furthermore we must also consider the potentials to which these branes couple, which we have not yet done. However, generalised charges can be constructed for these states if appropriate use of the isometries present in both cases are made. We will return to this problem in Sections \ref{sec:11-d KK monopole charge} and \ref{sec:M9-charge} but first explore the simpler cases of the charges for the branes in 10-dimensions.

\section{Massless IIA Supergravity}\label{sec:massless IIA supergravity}
The analysis for 11-dimensions carries through to the
IIA case. As is well known, the IIA theory can be determined from
Kaluza-Klein dimensional reduction of the 11-dimensional theory. We give the reduction rules in our conventions in Appendix \ref{sec:reducs}, but here take the more instructive route of carrying out the analysis purely from the IIA point of view. A partial analysis has already been carried out in \cite{Saffin:2004ar}, here we extend those results.
\subsection{Field equations of IIA Supergravity}
The IIA action in our conventions in the string frame is given by
\begin{eqnarray}
S_{IIA}=\frac{1}{2}\int
d^{10}x\sqrt{-g}\biggl[e^{-2\phi}\biggl(R+4|\nabla
\phi|^2-\frac{1}{2.3!}|H|^2\biggr)-\frac{1}{2}\sum_{n}\frac{1}{n!}|F^{(n)}|^2
\biggr]
\end{eqnarray}
where $n=2,4$, plus a Chern-Simons term:
\begin{eqnarray}
-\frac{1}{2}\int \frac{1}{2}dC^{(3)}\wedge dC^{(3)}\wedge B
\end{eqnarray}
Varying the action with respect to the potentials we get the (source free)
equations of motion for the field strengths, which together with the Bianchi identity for $F^{(2)}$, are:
\begin{eqnarray}
dF^{(2)}&=&0\\
dF^{(4)}&=&H\wedge F^{(2)}\\
dF^{(6)}&=&H\wedge F^{(4)}\\
dF^{(8)}&=&H\wedge F^{(6)}\\
dH^{(7)}&=&F^{(6)}\wedge F^{(2)}-\frac{1}{2}F^{(4)}\wedge F^{(4)}
\end{eqnarray}
where the dual field strengths have been defined as
\begin{eqnarray}
\ast F^{(2)}=F^{(8)} \qquad \ast F^{(4)}=-F^{(6)} \qquad
e^{-2\phi}\ast H=H^{(7)}
\end{eqnarray}
Consequently we can define our field potentials as follows
\begin{eqnarray}\label{eq:IIA dC1}
dC^{(1)}&=&F^{(2)}\\
dC^{(3)}&=&F^{(4)}+H\wedge C^{(1)}\\
dC^{(5)}&=&F^{(6)}+H\wedge C^{(3)}\\\label{eq:IIA dC7}
dC^{(7)}&=&F^{(8)}+H\wedge C^{(5)}\\\label{eq:IIA dC9}
dC^{(9)}&=&F^{(10)}+H\wedge C^{(7)}\\
dB&=&H\\ \label{eq:IIA dB6}
dB^{(6)}&=&H^{(7)}-C^{(1)}\wedge F^{(6)}+\frac{1}{2}C^{(3)}\wedge dC^{(3)}
\end{eqnarray}
where the number in parenthesis indicates the rank of the potential. The Ramond-Ramond potentials are denoted $C^{(n)}$ while the NS potentials by $B$ and $B^{(6)}$. Once again we have included the dual potentials, explicitly $C^{(5)}$ is the dual potential of $C^{(3)}$, $C^{(7)}$ is of $C^{(1)}$ and $B^{(6)}$ is of $B$. We omit the rank label on the 2-form $B$ for brevity. Note also that a 9-form potential and 10-form field strength have also been included. These appear in the massive IIA theory \cite{Bergshoeff:1996ui} where the field strength $F^{(10)}$ is the Hodge dual of the mass parameter, and the potential $C^{(9)}$ was introduced as a non-dynamical auxiliary potential in the action. We will consider this added complication in Section \ref{sec:massive IIA} but for now assume that we are in the massless theory with the mass parameter, and therefore $F^{(10)}$, set to zero. We do not neglect $C^{(9)}$ at this stage due to our democratic approach to the potentials and the fact that it is still related via T-duality to the other potentials.\footnote{Note that other potentials exist that are discussed in \cite{Bergshoeff:2006qw}. They can be determined by considering the full supersymmetry algebra and what possible potentials are allowed. They are not needed for our purposes here, but are allowed as they provide no additional propagating degrees of freedom.}

\subsection{Charges in IIA Supergravity}
Once again we can construct bilinear forms out of products of Killing spinors and the $\Gamma$ matrices. The bilinears that exist in IIA are found to be
\\
\\
\begin{tabular}[c]{c c c}
\nonumber $X=\overline{\epsilon}\Gamma_{11}\epsilon$ & $K_{\mu}=\overline{\epsilon}\Gamma_{\mu}\epsilon$& $\tilde{K}_{\mu}=\overline{\epsilon}\Gamma_{\mu} \Gamma_{11}\epsilon$\\
$\Omega_{\mu_1\mu_2}=\overline{\epsilon}\Gamma_{\mu_1\mu_2}\epsilon$ & $Z_{\mu_1\ldots \mu_4}=\overline{\epsilon}\Gamma_{\mu_1\ldots \mu_4} \Gamma_{11}\epsilon$&  $\Sigma_{\mu_1\ldots \mu_5}=\overline{\epsilon}\Gamma_{\mu_1\ldots \mu_5}\epsilon$
\end{tabular}
\\ \\We have defined our spinor here to be the sum of the chiral and anti-chiral spinors usually found in IIA
\begin{eqnarray}
\epsilon=\epsilon^++\epsilon^-
\end{eqnarray}
where
\begin{eqnarray}
\Gamma_{11}\epsilon^{\pm}=\Gamma_{012345679}\epsilon^{\pm}=\pm
\epsilon^{\pm}
\end{eqnarray}
The dual bilinears are defined as
\begin{eqnarray}
\tilde{\Sigma}_{(5)}=&\overline{\epsilon}\Gamma_{(5)} \Gamma_{11}\epsilon&=-\ast \Sigma_{(5)}\\
\Lambda_{(6)}=&\overline{\epsilon}\Gamma_{(6)}\epsilon&=\ast Z_{(4)}\\
\Psi_{(8)}=&\overline{\epsilon}\Gamma_{(8)} \Gamma_{11}\epsilon&=-\ast \Omega_{(2)}\\
\Pi_{(9)}=&\overline{\epsilon}\Gamma_{(9)}\epsilon&=-\ast \tilde{K}_{(1)}\\
\tilde{\Pi}_{(9)}=&\overline{\epsilon}\Gamma_{(9)} \Gamma_{11}\epsilon&=-\ast K_{(1)}\\
\Upsilon_{(10)}=&\overline{\epsilon}\Gamma_{(10)}\epsilon&=\ast X_{(0)}
\end{eqnarray}
where once again the ranks of the bilinears have been included for convenience.\\
\indent The spectrum of `standard' branes in IIA that can be supersymmetrically coupled to a potential consists of D0, D2, D4, D6 and D8-branes, and NS1(F-string), NS5 and NS9-branes \cite{Bergshoeff:2006qw}. The gravitational wave and KK monopole with 6-dimensional worldvolume are also 1/2-BPS states, but are purely gravitational. The flat target space SUSY algebra of IIA reads
\begin{eqnarray}\label{eq:IIA SUSY alg}
\nonumber \{Q_\alpha,Q_\beta\}&=&(C\Gamma^{\mu})_{\alpha
\beta}P_{\mu}+(C\Gamma_{11})_{\alpha
\beta}P_{11}+(C\Gamma^{\mu}\Gamma_{11})_{\alpha \beta}Z_{\mu}\\
\nonumber &&+\frac{1}{2}(C\Gamma ^{\mu_1 \mu_2})_{\alpha \beta}Z_{\mu_1
\mu_2}+\frac{1}{4!}(C\Gamma^{\mu_1\ldots \mu_4} \Gamma_{11})_{\alpha
\beta}Z_{\mu_1 \ldots \mu_4}\\ &&+\frac{1}{5!}(C\Gamma ^{\mu_1\ldots \mu_5 })_{\alpha \beta}Z_{\mu_1\ldots \mu_5}
\end{eqnarray}
where $C$ is the charge conjugation matrix. Once again there is a correspondence between the states and bilinears which we list in Table \ref{table:IIA states}.

\begin{table}[ht]
\centering
\begin{tabular}{|c||c|c|c|}
\hline 
BPS state & charge & bilinear & potential\\
\hline
D0 & $P_{11}$ & $X$ & $C^{(1)}$\\
\hline
D2 & $Z_{i_1i_2}$ & $\Omega$ & $C^{(3)}$\\
\hline
D4 & $Z_{i_1\ldots i_4}$ & $Z$ & $C^{(5)}$\\
\hline
D6 & $\ast(Z_{0i_1i_2i_3})$ & $\Lambda$ & $C^{(7)}$\\
\hline
D8 & $\ast(Z_{0i_1})$ & $\Psi$ & $C^{(9)}$\\
\hline
F-string & $Z_{i}$ & $\tilde{K}$ & $B$\\
\hline
NS5 & $Z_{i_1\ldots i_5}$ & $\Sigma$ & $B^{(6)}$\\
\hline
NS9 & $\ast (Z_{0})$ & $\Pi$ & $D^{(10)}$\\
\hline
KK monopole & $\ast (Z_{0i_1\ldots i_4})$ & $\tilde{\Sigma}$ & $i_{\alpha}N^{(7)}$\\
\hline
\end{tabular}
\caption{Branes, charges and their associated bilinears in the IIA theory. Also included are the potentials that minimally couple to the branes.}
\label{table:IIA states}
\end{table}

There are two independent Killing spinor equations in the bosonic IIA theory since there are two fermions which must have vanishing SUSY transformations. One is differential and the other algebraic, and are given by
\begin{eqnarray}\label{eq:IIA diff rel}
\nonumber \delta \psi_{\mu}&=&\nabla
_{\mu}\epsilon-\frac{1}{8}H_{\mu\nu_1\nu_2}\Gamma^{\nu_1\nu_2}\Gamma_{11}\epsilon\nonumber \\ &&-\frac{1}{8}\exp
(\phi)\biggl[\frac{1}{2}F^{(2)}_{\nu_1\nu_2}\Gamma^{\nu_1\nu_2}\Gamma_{\mu}\Gamma_{11}-\frac{1}{4!}
F^{(4)}_{\nu_1\ldots \nu_4}\Gamma^{\nu_1\ldots \nu_4}\Gamma_{\mu}\biggr]\epsilon\nonumber \\ &=&0\\\label{eq:IIA alg rel}
\nonumber \delta \lambda&=&\biggl[\partial_{\nu}\phi
\Gamma^{\nu}-\frac{1}{12}H_{\nu_1\nu_2\nu_3}\Gamma^{\nu_1\nu_2\nu_3}\Gamma_{11}\biggr]\epsilon\\ &&-\frac{1}{8}
\exp(\phi
)\biggl[3F^{(2)}_{\nu_1\nu_2}\Gamma^{\nu_1\nu_2}\Gamma_{11}-\frac{1}{12}F^{(4)}_{\nu_1\ldots \nu_4}\Gamma^
{\nu_1\ldots \nu_4}\biggr]\epsilon =0
\end{eqnarray}
respectively. Using the same method as in the 11-d case, differential relations for the bilinears can be obtained from (\ref{eq:IIA diff rel}). However these relations are not always simple and often involve messy contractions between terms. In order to simplify them one can obtain a set of algebraic relations by hitting (\ref{eq:IIA alg rel}) from the left with $\overline{\epsilon}\Gamma_{(p)}$ or $\overline{\epsilon}\Gamma_{(p)}\Gamma_{11}$, for various $p$. These algebraic relations can be substituted into the differential ones to yield reasonably simple looking equations.\footnote{An incomplete list of such relations was given in \cite{Saffin:2004ar}.} The details of such calculations are rather cumbersome and not of particular interest so only the results are presented here. Considering the bilinears corresponding to the D-brane charges first, we find
\begin{eqnarray}\label{eq:IIA dX}
d(e^{-\phi}X)&=&-i_KF^{(2)}\\\label{eq:IIA dOmega}
d(e^{-\phi}\Omega)&=&-e^{-\phi}XH+i_KF^{(4)}+\tilde{K}\wedge F^{(2)}\\
d(e^{-\phi}Z)&=&-e^{-\phi}\Omega\wedge H-i_KF^{(6)}-\tilde{K}\wedge
F^{(4)}\\\label{eq:IIA dLambda}
d(e^{-\phi}\Lambda)&=&-e^{-\phi}Z\wedge H+i_KF^{(8)}+\tilde{K}\wedge F^{(6)}\\\label{eq:IIA dPsi}
d(e^{-\phi}\Psi)&=&-e^{-\phi}\Lambda\wedge H-\tilde{K}\wedge F^{(8)}
\end{eqnarray}
as well as
\begin{eqnarray}\label{eq:IIA dKtilde}
d\tilde{K}&=&i_KH
\end{eqnarray}
Furthermore one can confirm that $K$, the bilinear associated with the momentum, is Killing as in the 11-d case.\\
\indent With these relations and the field strength equations at our disposal we can go about constructing expressions that correspond to the generalised charges. The method we used for finding such expressions was to start with a given relation from (\ref{eq:IIA dX}) to (\ref{eq:IIA dPsi}) and to then set about determining how the RHS could be written as an exact form. This can be done quite systematically by essentially listing all possible terms that could appear, differentiating them and then cancelling the resulting terms with the RHS of the differential bilinear relation under consideration.\\
\indent Ultimately the generalised charges $M_{(p)}$ for the D$p$-branes were found to be
\begin{eqnarray}\label{eq:IIA d0-charge}
M_{(0)}&=&e^{-\phi}X-i_KC^{(1)}\\
M_{(2)}&=&e^{-\phi}\Omega+i_KC^{(3)}+\tilde{K}\wedge C^{(1)}+M_{(0)}B\\
M_{(4)}&=&e^{-\phi}Z-i_KC^{(5)}-\tilde{K}\wedge
C^{(3)}+M_{(2)}\wedge
B-\frac{1}{2}M_{(0)}(B)^2\\
\nonumber M_{(6)}&=&e^{-\phi}\Lambda+i_KC^{(7)}+\tilde{K}\wedge
C^{(5)}+M_{(4)}\wedge B-\frac{1}{2}M_{(2)}\wedge (B)^2\\ &&+\frac{1}{3!}M_{(0)}(B)^3\\
\nonumber M_{(8)}&=&e^{-\phi}\Psi-i_KC^{(9)}-\tilde{K}\wedge
C^{(7)}+M_{(6)}\wedge B-\frac{1}{2}M_{(4)}\wedge (B)^2\\\label{eq:IIA d8-charge}
&&+\frac{1}{3!}M_{(2)}\wedge (B)^3-\frac{1}{4!}M_{(0)}(B)^4 
\end{eqnarray}
where gauges have been chosen such that ${\cal L}_K$ vanishes for each potential, as discussed in the 11-d case.\\
\indent In determining these charges it was found that such closed expressions only exist for on-shell field configurations. In other words when the potentials are defined arbitrarily so that the equations of motion for the field strengths are not necessarily satisfied it is not generally possible to find a set of terms that close. When the equations of motion are satisfied then the expressions above are uniquely determined up to an exact term which obviously has no ultimate effect once an integration over a closed cycle is performed. The fact that such charges only exist for on-shell configurations is not surprising since this is what we require for the configuration to be a solution of the theory. Other charges corresponding to `exotic' branes can also be found but require additional fields to be introduced. Such cases will be discussed in \cite{Callister:2007}.\\
\indent We can compare the generalised charges here with the D-brane SUSY algebras in flat backgrounds found in \cite{Hammer:1997ts}. We find that the $\Gamma$ matrices, present here in the definitions of the bilinears, appear in the same combinations. This is to be expected since it should be possible to produce these algebras from the generalised charges. However we now see that the $\Gamma$ matrix structure for the flat space algebra generalises to curved spaces. Here we also see how the target space potentials should generalise. These are set to zero for the flat space case and so are therefore not explicitly present, but in principle they would, at least partially, arise in the method presented in \cite{Hammer:1997ts} from the Wess-Zumino term when calculating the conjugate momentum for any given background.\\
\indent Since we have not formally considered the worldvolume fields in our analysis we cannot concretely determine the worldvolume field structure of the charges. However, we observe that the nested structure of the D-brane charges is the same as the appearance of the pullbacks of the target space potentials in the Wess-Zumino term of the worldvolume actions, which can be written in the form of a complex
\begin{eqnarray}\label{eq:WZ action}
S_{WZ}\sim\int C^{(n)}\wedge e^{(dV-B)}
\end{eqnarray}
where all the terms of the desired rank are understood to be present, and $V$ is the Born-Infeld 1-form and a pullback on $B$ is implied. The connection here is obvious since the SUSY algebra can be inferred from SUSY variations of the action. However, this term suggests that the worldvolume field $V$ and NS 2-form potential $B$ should appear in the worldvolume generalised charges in the combination $(dV-B)$. This can also be argued on the grounds of gauge invariance. Introducing the appropriate $dV$ terms into the above generalised charges does not effect their closure since $dV$ is trivially closed, and it will always form wedge products with other closed objects. The result would also reproduce the algebras found in \cite{Hammer:1997ts}.\\
\indent The interpretation of the generalised charges follows the same lines as that given in \cite{Hammer:1997ts}. Essentially we see that there is a nested structure of charges within charges, which can be related to the possibility of having branes within branes discussed in \cite{Douglas:1995bn}. This was present in the flat case but extends to curved space cases as we also found for the M5-brane. We have written the charges above in a fashion that explicitly shows this. The author of \cite{Hammer:1997ts} was unsure of how to interpret one of the terms which corresponds to the $\tilde{K}$ terms here. From comparing with (\ref{eq:IIA F-charge}) below, we see that there is in fact a natural interpretation involving string configurations within the branes. The fact that the entire string charge does not appear in any of the above D-brane charges, due to the lack of any $i_{K}B$ term, is not too problematic since it can be incorporated by performing a simple redefinition of the RR potentials, for example for the D2-brane charge we could make the redefinition $C^{(3)}\rightarrow C^{(3)}+C^{(1)}\wedge B$.\\
\indent The presence of the string coupling $g_s=e^{\phi}$ in the generalised charges occurs due to the presence of the brane tensions $T$ in the flatspace charges, (\ref{eq:top charge}) (recall that for the 1/2-BPS states loosely speaking we have $Q_{(p)}=T$.) Generally the tension can be a function of the scalars of the theory, in this case $g_s$. So for curved backgrounds the tension is no longer constant and therefore must be taken inside the integral. For D-branes the tension is proportional to $g_s^{-1}$ and so it is natural that we see this factor multiplying the leading bilinears in the above charges.\\
\indent We comment that the D8-brane is actually only a solution to massive IIA supergravity, \cite{Bergshoeff:1996ui}, so it might seem rather formal that we have considered its charge here for the massless IIA theory. If one were to actually calculate the charge in the massless case it would simply be zero. However, when we consider the massive IIA theory in Section \ref{sec:massive IIA}, we see that (\ref{eq:IIA d8-charge}) is in fact the appropriate expression, but with a different set of gauge conditions than the ones stated here. We have included the charge at this stage so that all the D-brane charges are listed together and the similarity in their structure can be seen.\\
\indent As a quick note, we mention that from using the reduction rules given in Appendix \ref{sec:reducs} it can easily be confirmed that the D2-charge corresponds to the direct dimensional reduction of the M2-charge (\ref{eq:11-d 2-charge}), while the D4-charge corresponds to the double dimensional reduction of the M5-charge (\ref{eq:11-d 5-charge}), which agrees with the reduction relations of the branes themselves as one would hope. Note also that the gauge conditions we have chosen for the potentials are also consistent under these reductions.\\
\indent We now consider the NS-branes. The relevant differential relations here are (\ref{eq:IIA dKtilde}) along with 
\begin{eqnarray}
\label{eq:IIA dSigma}
d(e^{-2\phi}\Sigma)&=&i_K H^{(7)}+e^{-\phi}\biggl[-Z\wedge
F^{(2)}-\Omega\wedge F^{(4)}-XF^{(6)}\biggr]\\
d(e^{-2\phi}\Pi)&=&0
\end{eqnarray}
The F-string and NS5-charge are then found to be
\begin{eqnarray}\label{eq:IIA F-charge}
{\cal M}_{(1)}&=&\tilde{K}+i_KB\\\label{eq:IIA NS5-charge}
\nonumber {\cal M}_{(5)}&=&e^{-2\phi}\Sigma+e^{-\phi}(Z\wedge
C^{(1)}+\Omega\wedge C^{(3)}+XC^{(5)})+i_KB^{(6)}\\ &&+\tilde{K}\wedge C^{(1)}\wedge
C^{(3)}-i_KC^{(1)}C^{(5)}+\frac{1}{2}i_KC^{(3)}\wedge C^{(3)}
\end{eqnarray}
respectively, where we have ${\cal L}_KB^{(6)}=0$ along with the previous gauge conditions . The F-string charge has a simple structure essentially due to its low rank. The NS5-brane charge on the other hand is considerably more complicated. The presence of the D-brane bilinears reflects the possibility of dissolving the lower rank D-branes within the NS5-brane. The full D-brane charges are implicit in (\ref{eq:IIA NS5-charge}) and can be made explicit by redefining the potentials. The essential structure is that of the leading bilinears of the lower rank D-brane charges. Note the factor of $g_s^{-2}$ which is present. This corresponds to the tension of the NS5-brane.\\
\indent From uplifting these charges to 11 dimensions it is found that the F-string charge is the double reduction of the M2-charge and the NS5-charge is the direct reduction of the M5-charge. This corresponds to the same relationships found for the branes themselves, as is required.\\
\indent Trivially the NS9-charge would seem to be given by
\begin{eqnarray}
\label{eq:IIA NS9-charge}
{\cal M}_{(9)}&=&e^{-2\phi}\Pi
\end{eqnarray}
However, we would not expect this to correspond to the full NS9-charge since when dealing with spacetime filling branes the high rank of the charge means that many terms become closed trivially. For instance from \cite{Bergshoeff:2006qw} we see that the Wess-Zumino term in the action is simply 
\begin{eqnarray}
\int_{NS9}D^{(10)}
\end{eqnarray}
where $D^{(10)}$ is the non-dynamical 10-from potential to which the NS9-brane couples. We would therefore expect a term of the form $i_KD^{(10)}$ to appear in the NS9-charge, in analogy to the other branes. Such a term would be closed trivially (assuming ${\cal L}_KD^{(10)}=0$) and so could be added to (\ref{eq:IIA NS9-charge}) and the resulting expression would still be closed. In other words, simply demanding that an expression be closed leads to ambiguities when finding the charge for spacetime filling branes, as is the case here. We should also not forget that it is actually inconsistent to consider a single spacetime filling brane on its own due to gauge anomalies and charge non-conservation, \cite{Polchinski:1995mt,Bergshoeff:1998re}. To make things consistent one has to have 32 such branes and carry out some $N=1$ truncation, the details of which will not concern us here but see \cite{Bergshoeff:1998re}. This may need to be considered when determining the expression for the charge.\\
\indent Looking at (\ref{eq:IIA NS9-charge}) we do notice however that the appearance of the string coupling matches that found in the brane tension according to \cite{Bergshoeff:2006qw}. However other references on the subject (for example \cite{Bergshoeff:1998re}) state that the this factor should be $e^{-4\phi}$. This discrepancy presumably arises because in the latter case the 9-brane arises from a direct reduction of the M9-brane and so contains a gauged isometry in its worldvolume. This issue is discussed somewhat in the forthcoming article \cite{Callister:2007}. Furthermore, it is also discussed in \cite{Bergshoeff:2006qw} that there are actually 2 spacetime-filling branes in IIA. However only the NS9-brane supersymmetrically couples to a potential and so we would perhaps only expect to find a charge for that brane.\\ 
\indent Finally we are left with the following differential relation to consider for $\tilde{\Sigma}$:
\begin{eqnarray}\label{eq:IIA sigmatilde dif}
\nonumber d\tilde{\Sigma}_{a_1\ldots a_6}&=&6e^{\phi}F^{(2)}_{b[a_1}\Lambda_{a_2\ldots a_6]}^{\phantom{a_2\ldots a_6]}b}-e^{\phi}F^{(4)}_{b_1\ldots b_3[a_1}\Psi_{a_2\ldots a_6]}^{\phantom{a_2\ldots a_6]}b_1\ldots b_3}\\
&&+15H_{b[a_1a_2}\Sigma_{a_3\ldots a_6]}^{\phantom{a_3\ldots a_6]}b}
\end{eqnarray}
together with an algebraic relation obtained by hitting (\ref{eq:IIA alg rel}) in an orthonormal frame with $\Gamma_{a_1\ldots a_6}\Gamma_{11}$: 
\begin{eqnarray}\label{eq:IIA sigmatilde alg}
0&=&(d\phi\wedge \tilde{\Sigma})_{a_1\ldots a_6}+\frac{1}{2}(i_H\Pi)_{a_1\ldots a_6}-\frac{15}{2}H_{b[a_1a_2}\Sigma_{a_3\ldots a_6]}^{\phantom{a_3\ldots a_6]}b}\\\nonumber &&+e^{\phi}\biggl[-\frac{9}{2}F^{(2)}_{b[a_1}\Lambda_{a_2\ldots a_6]}^{\phantom{a_2\ldots a_6]}b}+\frac{1}{4}F^{(4)}_{b_1\ldots b_3[a_1}\Psi_{a_2\ldots a_6]}^{\phantom{a_2\ldots a_6]}b_1\ldots b_3}\\ &&-5F^{(4)}_{b[a_1\ldots a_3}Z_{a_4\ldots a_6]}^{\phantom{a_4\ldots a_6]}b}\biggr]
\end{eqnarray}
which we expect should be used in finding a charge for the KK monopole. There is a problem here which is essentially the same as what was found in the eleven dimensional case for (\ref{eq:11-d dLambda}). The problem is that even combining these relations does not produce a differential relation for $\tilde{\Sigma}$ which is free from the problematic contractions on the RHS. Thus they alone cannot be used to determine the structure of a 6-form closed expression that would correspond to the charge of the IIA KK monopole. The solution to this is to incorporate the isometry which exists in the spacetime solution of the KK monopole, into the charge as for the 11-d case. We carry out this procedure in Section \ref{sec:10-d KK-charges}. 

\section{Type IIB  Supergravity}\label{sec:type IIB supergravity}

\subsection{Field equations  in IIB Supergravity}

The IIB case follows on analogously from the others. The standard string frame IIB action that is used in the literature reads as:
\begin{eqnarray}
S_{IIB}&=&\frac{1}{2}\int
d^{10}x\sqrt{-g}\biggl[e^{-2\varphi}\biggl(R+4|\nabla
\varphi|^2-\frac{1}{2.3!}|{\cal H}|^2_{(3)}\biggr)\nonumber \\ &&-\frac{1}{2}\sum_{n=1,3}\frac{1}{n!}|{\cal F}^{(n)}|^2-\frac{1}{4.5!}|{\cal F}^{(5)}|^2\biggr]
\end{eqnarray}
with Chern-Simons term
\begin{eqnarray}
-\frac{1}{2}\int {\cal C}^{(4)}\wedge {\cal H}\wedge {\cal F}^{(3)}
\end{eqnarray}
The equations of motion or modified Bianchi identities for the field strengths are found to be
\begin{eqnarray}
d{\cal F}^{(1)}&=&0\\
d{\cal F}^{(3)}&=&{\cal H}\wedge {\cal F}^{(1)}\\
d{\cal F}^{(5)}&=&{\cal H}\wedge {\cal F}^{(3)}\\
d{\cal F}^{(7)}&=&{\cal H}\wedge {\cal F}^{(5)}\\
d{\cal F}^{(9)}&=&{\cal H}\wedge {\cal F}^{(7)}\\
d{\cal H}^{(7)}&=&{\cal F}^{(1)}\wedge {\cal F}^{(7)}-{\cal F}^{(3)}\wedge {\cal F}^{(5)}
\end{eqnarray}
where the dual field strengths have been defined as
\begin{eqnarray}
\ast {\cal F}^{(1)}=-{\cal F}^{(9)} \qquad \ast {\cal F}^{(3)}={\cal F}^{(7)} \qquad e^{-2\varphi}\ast {\cal H}={\cal H}^{(7)}
\end{eqnarray}with ${\cal F}^{(5)}$ being anti self-dual with our conventions for the Hodge dual operator. As in the previous case the numbers in parenthesis indicate the rank of the fields, with the exception of ${\cal H}$, the 3-form NS field strength, where the rank is omitted for brevity.\\
\indent  We then define the IIB potentials as
\begin{eqnarray}
d{\cal C}^{(0)}&=&{\cal F}^{(1)}\\
d{\cal C}^{(2)}&=&{\cal F}^{(3)}+{\cal C}^{(0)}{\cal H}\\
d{\cal C}^{(4)}&=&{\cal F}^{(5)}+{\cal C}^{(2)}\wedge {\cal H}\\
d{\cal C}^{(6)}&=&{\cal F}^{(7)}+{\cal C}^{(4)}\wedge {\cal H}\\
d{\cal C}^{(8)}&=&{\cal F}^{(9)}+{\cal C}^{(6)}\wedge {\cal H}\\
d{\cal C}^{(10)}&=&0\\
d{\cal B}&=&{\cal H}\\
d{\cal B}^{(6)}&=&{\cal H}^{(7)}+{\cal C}^{(2)}\wedge {\cal F}^{(5)}+\frac{1}{2}{\cal C}^{(2)}\wedge
{\cal C}^{(2)}\wedge {\cal H} - {\cal C}^{(0)}{\cal F}^{(7)}
\end{eqnarray}
where similar conventions hold as in the IIA case. We once again include the dual potentials since we are considering all branes equally. Note that we have formally included a ${\cal C}^{(10)}$ potential whose field strength will trivially be zero. Such a potential must exist because of T-duality, and is known to couple to the D9-branes \cite{Bergshoeff:2005ac,Bergshoeff:2006ic,Bergshoeff:1999bx} and is required for its action to be kappa-symmetric.\footnote{As in the IIA case, more potentials than this can be defined, however they generally are not required for our purposes, though we do briefly consider a 10-form potential ${\cal B}^{(10)}$ which should presumably appear in one of our 9-form charges. See \cite{Bergshoeff:2005ac,Bergshoeff:2006ic} for more details.}

\subsection{Charges in IIB Supergravity}
We begin our analysis of the brane charges by constructing the various bilinear forms. There are two 16 component spinors in the IIB theory
(\(\epsilon^1,\epsilon^2\)), both of which are chiral, i.e
\begin{eqnarray}
\Gamma_{11}\epsilon^i=\Gamma_{0123456789}\epsilon^i=\epsilon^i
\end{eqnarray}
The only non-zero bilinears that exist are found to be
\begin{eqnarray}
\nonumber K^{ij}_{\mu}&=&\epsilon^i\Gamma_{\mu} \epsilon^j\\
\Phi^{ij}_{\mu_1\mu_2\mu_3}&=&\epsilon^i\Gamma_{\mu_1\mu_2\mu_3} \epsilon^j\qquad i\neq j\\
\Sigma^{ij}_{\mu_1\ldots \mu_5}&=&\epsilon^i\Gamma_{\mu_1\ldots \mu_5} \epsilon^j
\end{eqnarray}
which are not all independent but satisfy the relations
\begin{eqnarray}
K^{12}=K^{21} \qquad \Phi^{12}=-\Phi^{21} \qquad
\Sigma^{12}=\Sigma^{21}
\end{eqnarray}
For convenience we define
\begin{eqnarray}
K^{+}=\frac{1}{2}(K^{11}+K^{22}) \qquad
K^{-}=\frac{1}{2}(K^{11}-K^{22})\\
\Sigma^{+}=\frac{1}{2}(\Sigma^{11}+\Sigma^{22}) \qquad
\Sigma^{-}=\frac{1}{2}(\Sigma^{11}-\Sigma^{22})
\end{eqnarray}
We again also consider the
dual bilinears which are given as:
\begin{eqnarray}
\Pi^{ij}_{(7)}=&\overline{\epsilon}^i\Gamma_{(7)}\epsilon^j&=+\ast \Phi_{(3)}^{ij}\\
\Omega^{ij}_{(9)}=&\overline{\epsilon}^i\Gamma_{(9)}\epsilon^j&=-\ast K_{(1)}^{ij}
\end{eqnarray}
with each $\Sigma^{ij}$ being anti self-dual in our conventions.\\
\indent The standard brane spectrum of the IIB theory consists of D1,
D3, D5, D7 and D9-branes, and F-strings, NS5 and NS9-branes. The gravitational wave and the KK monopole with a 6-dimensional worldvolume are also BPS states but are purely gravitational as in the IIA case. The full SUSY algebra reads as
\begin{eqnarray}\label{eq:IIB SUSY alg}
\nonumber \{Q^i_\alpha,Q^j_\beta\}&=&(C{\cal P}^+\Gamma^{m})_{\alpha
\beta}(\delta^{ij}P_{\mu}+\sigma_3^{ij}Z_{\mu}+\sigma_1^{ij}\tilde{Z}_{\mu})\\
\nonumber &&+\frac{1}{3!}i\sigma_2^{ij}(C{\cal P}^+\Gamma^{\mu_1\mu_2\mu_3})_{\alpha
\beta}Z_{\mu_1\mu_2\mu_3}+\frac{1}{5!}\delta^{ij}(C{\cal P}^+\Gamma^{\mu_1\ldots \mu_5})_{\alpha
\beta}Z_{\mu_1\ldots \mu_5}\\ &&+\frac{1}{5!}(C{\cal P}^+\Gamma^{\mu_1\ldots \mu_5})_{\alpha
\beta}(\sigma_3^{ij}\tilde{Z}_{\mu_1\ldots \mu_5}+\sigma_1^{ij}\overline{Z}_{\mu_1\ldots \mu_5})
\end{eqnarray}
where ${\cal P}^+=\frac{1}{2}(1+\Gamma_{11})$ is a chiral projector, $\sigma_i$ are the Pauli matrices and the 5-form charges are anti self-dual. The relations between the 1/2-BPS states, charges and bilinears are summarised in Table \ref{table:IIB states}. Here the duals of the 5-form charges are not considered because they are not independent of the original charges. The correspondence between the branes and the potentials is given in \cite{Bergshoeff:2006ic} for example.\\

\begin{table}[ht]
\centering
\begin{tabular}{|c||c|c|c|}
\hline 
BPS state & charge & bilinear & potential\\
\hline
D1 & $\tilde{Z}_{i}$ & $K^{12}$ & ${\cal C}^{(2)}$\\
\hline
D3 & $Z_{i_1i_3i_2}$ & $\Phi^{12}$ & ${\cal C}^{(4)}$\\
\hline
D5 & $\overline{Z}_{i_1\ldots i_5}$ & $\Sigma^{12}$ & ${\cal C}^{(6)}$\\
\hline
D7 & $\ast(Z_{0i_1i_2})$ & $\Pi^{12}$ & ${\cal C}^{(8)}$\\
\hline
D9 & $\ast(\tilde{Z}_{0})$ & $\Omega^{12}$ & ${\cal C}^{(10)}$\\
\hline
F-string & $Z_{i}$ & $K^{-}$ & ${\cal B}$\\
\hline
NS5 & $\tilde{Z}_{i_1\ldots i_5}$ & $\Sigma^{-}$ & ${\cal B}^{(6)}$\\
\hline
NS9 & $\ast (Z_{0})$ & $\Omega^{-}$ & ${\cal B}^{(10)}$\\
\hline
KK monopole & $Z_{i_1\ldots i_5}$ & $\Sigma^{+}$ & $i_{\alpha}{\cal N}^{(7)}$\\
\hline
\end{tabular}
\caption{Branes, charges and their associated bilinears in the IIB theory. Also included are the potentials that minimally couple to the branes.}
\label{table:IIB states}
\end{table}

Parallel to the IIA case we find two Killing spinor equations in the IIB theory, one differential and the other algebraic given respectively by
\begin{eqnarray}
D_{\mu}\epsilon&=&0\\
{\cal P}\epsilon&=&0
\end{eqnarray}
where
\begin{eqnarray}
\nonumber D_{\mu}&=&\nabla_{\mu}+\frac{1}{8}H_{\mu \nu_1\nu_2}\Gamma^{\nu_1\nu_2}\otimes
\sigma_3\\ &&+\frac{1}{16}e^{\varphi}\sum_{n=1}^{5}\frac{(-1)^{n-1}}{(2n-1)!}{\cal F}^{(2n-1)}_{\nu_1\ldots \nu_{2n-1}}\Gamma^{\nu_1\ldots \nu_{2n-1}}\Gamma_{\mu}\otimes
\lambda_n\\
\nonumber
{\cal
P}&=&\Gamma^{\nu}\partial_{\nu}\varphi+\frac{1}{12}H_{\nu_1\ldots \nu_3}\Gamma^{\nu_1\ldots \nu_3}\otimes
\sigma_3\\ &&+\frac{e^{\varphi}}{4}\sum_{n=1}^5\frac{(-1)^{n-1}(n-3)}{(2n-1)!}{\cal F}^{(2n-
1)}_{\nu_1\ldots \nu_{2n-1}}\Gamma^{\nu_1\ldots \nu_{2n-1}}\otimes
\lambda_n
\end{eqnarray}
and
\begin{displaymath}
\lambda_n=\left\{ \begin{array}{c c}
\sigma_1 &\textrm{if n even}\\
i\sigma_2 &\textrm{if n odd}
\end{array} \right.
\end{displaymath}
Following the same procedure as for the IIA case we can produce a set of differential relations for the bilinears. Those relevant to the D-branes are\footnote{The unsimplified relations can be found in \cite{Hackett-Jones:2004yi}.}
\begin{eqnarray}
d(e^{-\varphi}K^{12})&=& -K^{-}\wedge {\cal F}^{(1)}+i_{K^{+}}{\cal F}^{(3)}\\
d(e^{-\varphi}\Phi^{12})&=&-e^{-\varphi}H\wedge K^{12}+K^{-}\wedge
{\cal F}^{(3)}-i_{K^{+}}{\cal F}^{(5)}\\
d(e^{-\varphi}\Sigma^{12})&=&-e^{-\varphi}H\wedge \Phi^{12}-K^{-}\wedge
{\cal F}^{(5)}+i_{K^{+}}{\cal F}^{(7)}\\
d(e^{-\varphi}\Pi^{12})&=&-e^{-\varphi}H\wedge \Sigma^{12}+K^{-}\wedge
{\cal F}^{(7)}-i_{K^{+}}{\cal F}^{(9)}\\
d(e^{-\varphi}\Omega^{12})&=&-e^{-\varphi}H\wedge\Pi^{12}-K^{-}\wedge {\cal F}^{(9)}
\end{eqnarray}
and also
\begin{eqnarray}\label{eq:IIB dK44}
dK^{-}&=&-i_{K^{+}}{\cal H}
\end{eqnarray}
Furthermore it can be shown that $K^{+}$ is Killing.\\
\indent With these relations in hand we can determine the structure of the generalised charges $N_{(p)}$ for the IIB D$p$-branes. They are found to be
\begin{eqnarray}\label{eq:IIA d1-charge}
N_{(1)}&=&e^{-\varphi}K^{12}-{\cal C}^{(0)}K^{-}+i_{K^{+}}{\cal C}^{(2)}\\
N_{(3)}&=&e^{-\varphi}\Phi^{12}+K^{-}\wedge
{\cal C}^{(2)}-i_{K^{+}}{\cal C}^{(4)}+N_{(1)}\wedge B\\
\nonumber N_{(5)}&=&e^{-\varphi}\Sigma^{12}-K^{-}\wedge
{\cal C}^{(4)}+i_{K^{+}}{\cal C}^{(6)}+N_{(3)}\wedge B\\
&&-\frac{1}{2}N_{(1)}\wedge (B)^2\\
\nonumber N_{(7)}&=&e^{-\varphi}\Pi^{12}+K^{-}\wedge
{\cal C}^{(6)}-i_{K^{+}}{\cal C}^{(8)}+N_{(5)}\wedge B\\
&&-\frac{1}{2}N_{(3)}\wedge (B)^2+\frac{1}{3!}N_{(1)}\wedge
(B)^3\\
\nonumber N_{(9)}&=&e^{-\varphi}\Omega^{12}-K^{-}\wedge
{\cal C}^{(8)}+i_{K^{+}}{\cal C}^{(10)}+N_{(7)}\wedge B\\\nonumber
&&-\frac{1}{2}N_{(5)}\wedge (B)^2+\frac{1}{3!}N_{(3)}\wedge
(B)^3\\ &&-\frac{1}{4!}N_{(1)}\wedge (B)^4
\end{eqnarray}
where ${\cal L}_{K^{+}}$ of all the potentials vanish.\\
\indent The structure of the charges is essentially the same as for the IIA D-branes, and the comments made in that instance generally apply to the IIB case as well. However, a few additional comments are in order. Although generally the tensions factors here are the same as for the IIA case, there is a discrepancy for the D1-brane which has a tension factor of $\sqrt{e^{-2\varphi}+({\cal C}^{(0)})^2}$, \cite{Bergshoeff:2006ic}. Re-examining generalised charge for the D1-brane (\ref{eq:IIA d1-charge}) we see that it is unique amongst the charges we have constructed so far in that it actually contains two bilinears that are of the same rank as the charge itself. So although for flat backgrounds it is $K^{12}$ that is the bilinear that would form the charge, in some sense for general backgrounds both $K^{12}$ and $K^+$ are on a more equal footing.\\
\indent Thinking of the charge as a generalised calibration, both these bilinears, when pulled back to a 1 dimensional line, should obey some bound related to the length of the line, e.g. $K^{12}|_{line}\leq length$. Furthermore, the bilinears are orthogonal in the sense that for a line where one of the bilinears satisfies the bound, the other will vanish. We would then find a bound of the form 
\begin{eqnarray}
(e^{-\varphi}K^{12}-{\cal C}^{(0)}K^{-})|_{line} \leq length \sqrt{e^{-2\varphi}+({\cal C}^{(0)})^2} \end{eqnarray}
For BPS states this would then give the condition that
\begin{eqnarray}
T=\sqrt{e^{-2\varphi}+({\cal C}^{(0)})^2}
\end{eqnarray}
which is in agreement with the tension factor for the D1-brane. A similar effect will happen for $(p,q)$-string bound states.\\
\indent Another point here for the IIB case is that we have assumed that our 7-form charge here corresponds to the D7-brane. This seems natural as it has a similar structure to the other D-brane charges. However, the subject of 7-branes in the IIB theory has some technicalities, \cite{Meessen:1998qm,Bergshoeff:2006ic,Eyras:1999at,Bergshoeff:2006jj}. The 8-form potential ${\cal C}^{(8)}$, to which the D7-brane couples, transforms as part of a triplet under $SL(2,\mathbb{Z})$. Other charges for the branes that couple to these other potentials will also exist and also involve the 7-form bilinear\footnote{This is because all the 7-branes correspond to the same charge in the flatspace SUSY algebra, \cite{Bergshoeff:2006ic}.} $\Pi^{12}$ but exhibit different `tension factors'. In order to find these charges field equations for these other potentials must be defined. Work on this subject is currently underway and will be presented in \cite{Callister:2007}.\\
\indent The situation with the 9-branes is similar to that of the 7-branes except in this instance there are thought to be six branes in total, \cite{Bergshoeff:2006ic}. Two of them transform as a doublet under $SL(2,\mathbb{Z})$ whereas the remaining four transform as a quadruplet. We do not explore all these possible branes due to the difficulties with spacetime filling branes already discussed for the IIA NS9-brane. We note however that it is natural to associate the 9-form charge above with the D9-brane since it follows the same structure as the other D-branes. We have included the term $i_{K^{+}}{\cal C}^{(10)}$ for this reason, even though it is not required to make the charge close. Its inclusion does seem to be required from T-duality however, which will be discussed in Section \ref{sec:T-dual}.\\
\indent Next we consider the NS-charges for which we require relation (\ref{eq:IIB dK44}) as well as
\begin{eqnarray}\label{eq:IIB dSigma}
\nonumber d(e^{-2\varphi}\Sigma^{-})&=&-i_{K^{+}}H^{(7)}+e^{-\varphi}\biggl[K^{12}\wedge
{\cal F}^{(5)}+\Phi^{12}\wedge {\cal F}^{(3)}\\ &&+\Sigma^{12}\wedge {\cal F}^{(1)}\biggr]\\
d(e^{-2\varphi}\Omega^{-})&=&0
\end{eqnarray} 
One then determines the charges for the F-string and NS5-brane to be
\begin{eqnarray}
{\cal N}_{(1)}&=&K^{-}-i_{K^{+}}B\\\label{eq:IIB NS5-charge}
\nonumber {\cal N}_{(5)}&=&e^{-2\varphi}\Sigma^{-}+e^{-\varphi}({\cal C}^{(0)}\Sigma^{12}+\Phi^{12}\wedge {\cal C}^{(2)}+K^{12}\wedge {\cal C}^{(4)})\\ \nonumber &&-i_{K^+}{\cal B}^{(6)}+\frac{1}{2}K^{-}\wedge
{\cal C}^{(2)}\wedge {\cal C}^{(2)}-{\cal C}^{(0)}K^{-}\wedge {\cal C}^{(4)}\\ &&+i_{K^{+}}{\cal C}^{(2)}\wedge
{\cal C}^{(4)}
\end{eqnarray}
where we have used the condition ${\cal L}_{K^{+}}{\cal B}^{(6)}=0$ in addition to our previous gauge choices. The F-string charge has essentially the same simplistic structure as for the IIA case. The  NS5-brane charge also shows similarity with the IIA case in that it contains all the leading bilinears of the lower rank D-brane charges. The tension of the NS5-brane has the same complication that was found for the D1-brane, and is given by $e^{-\varphi}\sqrt{e^{-2\varphi}+({\cal C}^{(0)})^2}$, see \cite{Bergshoeff:2006ic}. From the NS5-charge this is can be seen from the presence of two 5-form bilinears, $\Sigma^-$ and $\Sigma^{12}$. Using the same argument as given for the D1-brane case and considering the NS5-brane tension, the scalars multiplying these bilinears are what we would expect. Once again a similar effect will happen for $(p,q)$-5-brane bound states.\\
\indent We also find a 9-form charge to be trivially given by
\begin{eqnarray}\label{eq:IIB NS9-charge}
{\cal N}_{(9)}&=&e^{-2\varphi}\Omega^{-}
\end{eqnarray}
This charge presumably corresponds to the NS9-brane (called simply the `S9-brane' in \cite{Bergshoeff:2006ic}) since it has the expected dilaton dependence.  As in the IIA case we would not expect (\ref{eq:IIB NS9-charge}) to represent the full charge of the NS9-brane since one could add to it a term of the form $i_{K^{+}}{\cal B}^{(10)}$, where ${\cal B}^{(10)}$ is the potential which the NS9-brane couples to, and it would still remain closed, as well as perhaps other terms that close trivially due to their rank.\\
\indent Finally we find the following differential and algebraic relations for $\Sigma^{+}$ which we expect to be related to the KK monopole charge
\begin{eqnarray}
d\Sigma^{+}_{a_1\ldots a_6}&=&-15H^{b}_{\phantom{a}[a_1a_2}\Sigma^{-}_{a_3\ldots a_6]b}+e^{\varphi}\biggl[-\frac{3}{2}i_{{\cal F}^{(1)}}\Pi^{12}_{a_1\ldots a_6}\nonumber \\ && +\frac{3}{2}i_{{\cal F}^{(3)}}\Omega^{12}_{a_1\ldots a_6}-\frac{15}{2}{\cal F}^{(3)}_{b[a_1a_2}\Sigma^{12\phantom{\ldots \ a_6}b}_{a_3\ldots a_6]}\nonumber \\ && +\frac{15}{2}\Phi^{12}_{b[a_1a_2}{\cal F}^{(5)\phantom{\ldots a_6}b}_{a_3\ldots a_6]}\biggr]\\
0&=&\Sigma^{+}\wedge
d\varphi_{a_1\ldots a_6}+\frac{1}{2}i_H\Omega^{-}_{a_1\ldots a_6}-\frac{15}{2}H^{b}_{\phantom{a}[a_1a_2}\Sigma^
{-}_{a_3\ldots a_6]b}\nonumber \\ &&+e^{\varphi}\biggl[-i_{{\cal F}^{(1)}}\Pi^{12}_{a_1\ldots a_6}+\frac{1}{2}i_{{\cal F}^
{(3)}}\Omega^{12}_{a_1\ldots a_6}\nonumber \\ &&-\frac{15}{2}{\cal F}^{(3)}_{b[a_1a_2}\Sigma^{12\phantom{\ldots \ a_6}b}_{a_3\ldots a_6]}
\biggr]
\end{eqnarray}
As with the IIA and 11-d case, these relations alone cannot be used to determine the structure of the KK monopole charge. We present the actual charge in Section \ref{sec:10-d KK-charges}.

\section{Kaluza-Klein Monopole charges}\label{sec:Kaluza-Klein monopole charges}
In Section \ref{sec:11-d charges}  we constructed the generalised charges for the M2 and M5-branes but observed that there were difficulties in directly constructing similar charges for the KK monopole and M9-brane owing to the contractions present in the differential relation for $\hat{\Lambda}$, (\ref{eq:11-d dLambda}). In this section we address this problem by showing that these contractions can be removed for backgrounds exhibiting an isometry, and find the generalised charge for the KK monopole. We begin by considering the 11 dimensional case where we carry out this process explicitly. The method in the 10 dimensional cases is much the same so we simply quote the necessary results in those instances. The M9-brane will be considered in Section \ref{sec:M9-charge} where we discuss the massive supergravity theories.

\subsection{11 dimensional  KK-monopole charge}\label{sec:11-d KK monopole charge}
The KK-monopole \cite{Sorkin:1983ns,Gross:1983hb,Bergshoeff:1997gy,Eyras:1998hn}, in 11 dimensions, consists of a 7-dimensional worldvolume and a 4-dimensional transverse Taub-NUT space which contains a compact isometry.  It is the presence of this isometry that allows for the removal of the contractions from (\ref{eq:11-d dLambda}). \\
\indent The differential relations (\ref{eq:11-d dK}) - (\ref{eq:11-d dUpsilon}) were calculated using the Killing spinor equation (\ref{eq:11-d killingspinor}) which is valid in general bosonic supersymmetric backgrounds. However, when we are dealing with spacetimes with specific features such as the KK monopole background, these equations will not allude to any information arising from these specific features, since this would not apply generally. In other words it seems that it might be possible to find additional relations in these circumstances. For the case at hand where there is an isometry present, this is a fairly straight forward task. Simply put, as well as considering all the components of (\ref{eq:11-d killingspinor}) collectively, as was done previously, one can also consider the component along the isometry in isolation. In this instance the covariant derivative can be decomposed into its partial derivative and connection components. Since the partial derivative vanishes, we end up with an 11-d algebraic Killing spinor equation. This can then be used to create additional algebraic relations which can be combined with the differential relations in much the same way as was done in the 10-dimensional cases.\footnote{Actually this process is completely analogous to the 10-d cases, as the 11-d theory with an isometry is effectively just the IIA theory with a different interpretation. The new 11-d algebraic Killing spinor equation is just the 11-d version of the algebraic Killing spinor equation of IIA, and reduces to it upon compactification along the isometry.}   We now demonstrate this explicitly. In the following, we represent the isometry by the Killing vector $\hat{\alpha}$ and use the standard co-ordinate system where $\hat{\alpha}^{\mu}=\delta^{\mu z}$. We then find (\ref{eq:11-d killingspinor}) along the isometry reads as
\begin{eqnarray}\label{eq:11-d alg rel}
\nonumber \hat{\tilde{D}}_z\hat{\epsilon}&=&\hat{\nabla}_z\epsilon+\frac{1}{288}\hat{e}^{\underline{z}}_z\hat{\Gamma}_{\underline{z}}^{\phantom{z}abcd}\hat{F}_{abcd}\hat{\epsilon}-\frac{1}{36}\hat{e}^{\underline{z}}_z\hat{\Gamma}^{abcd}\hat{F}_{\underline{z}abc}\hat{\epsilon}\\
\nonumber&=&\biggl[-\frac{1}{8}d\hat{\alpha}_{ab}\hat{\Gamma}^{ab}-\frac{1}{4}\hat{R}^{-1}\partial_a(\hat{R}^2)\hat{\Gamma}^{a\underline{z}}\biggr]\hat{\epsilon}\\
&&+\biggl[\frac{1}{288}\hat{R}\hat{\Gamma}_{\underline{z}}^{\phantom{z}abcd}\hat{F}_{abcd}-\frac{1}{36}\hat{R}\hat{\Gamma}^{abc}\hat{F}_{\underline{z}abc}\biggr]\hat{\epsilon}=0
\end{eqnarray}    
where we have made use of the standard orthonormal basis which is stated in Appendix \ref{sec:conventions}. In this section the orthonormal indices are understood to not run over $\underline{z}$ since we are considering the isometry direction separately. Also note that $\hat{\alpha}_z=\hat{R}^2$ where $\hat{R}$ is the radius of the compact direction.\\
\indent Algebraic conditions of all ranks can be obtained from (\ref{eq:11-d alg rel}) but for the KK monopole the required condition is obtained by hitting it from the left with $\hat{\overline{\epsilon}}\hat{\Gamma}_{a_1\ldots a_7\underline{z}}$. The equation produced looks a bit messy but various terms can be expressed more tidily in terms of the Hodge duals of their constituent parts. After multiplying by a factor of $-4\hat{R}^{-1}$ one obtains:
\begin{eqnarray}\label{eq:11-d KK alg}
\nonumber 0&=&\biggl(\hat{R}^{-2}i_{\hat{K}}(i_{\hat{\alpha}}\hat{G}^{(9)})+\hat{R}^{-2}d\hat{\alpha}\wedge i_{\hat{\alpha}}\hat{\Lambda}-\hat{R}^{-2}d(\hat{R}^2)\wedge \hat{\Lambda}\\\nonumber &&+\frac{1}{3}\hat{R}^{-2}i_{\hat{\alpha}}\hat{\omega}\wedge i_{\hat{\alpha}}\hat{F}^{(7)}\biggr)_{a_1\ldots a_7}-\frac{14}{3}\hat{\omega}^a_{\phantom{a}[a_1}\hat{F}^{(7)}_{a_2\ldots a_7]a}\\ &&+\frac{35}{3}\hat{\Sigma}_{a[a_1\ldots a_4}\hat{F}_{a_5\ldots a_7]}^{\phantom{a_5\ldots a_7}a}+\frac{2}{3}\hat{R}^{-2}(i_{\hat{\alpha}}\hat{\Sigma}\wedge i_{\hat{\alpha}}\hat{F})_{a_1\ldots a_7}
\end{eqnarray}
Here we consider $\hat{\alpha}$ as a potential to a field strength $d\hat{\alpha}=\hat{G}^{(2)}$ as done explicitly in \cite{Sato:1999bu,Sato:2000mw}.  So $\hat{G}^{(9)}=\hat{\ast} \hat{G}^{(2)}$ is the 9-form dual field strength, with potential $\hat{N}^{(8)}$,  originally introduced in \cite{Bergshoeff:1997ak}. It is this potential to which the monopole couples.\\
\indent Once again note that none of the indices run over $\underline{z}$ including $a_1$ to $a_7$ due to the way they were introduced. The implication of this is that equation (\ref{eq:11-d KK alg})  is not valid if one of the free indices is replaced by $\underline{z}$, i.e. it is an equation whose $\underline{z}$ components are zero, but it is expressed in terms of objects that have non-zero $\underline{z}$ components, and therefore care must be taken when changing from one basis to another.\\
\indent Now considering (\ref{eq:11-d dLambda}) in the orthonormal basis and only those components not containing $\underline{z}$, we can add (\ref{eq:11-d KK alg}) to the RHS to obtain:
\begin{eqnarray}\label{eq:11-d dLambda 2}
\nonumber \hat{R}^2d\hat{\Lambda}_{a_1\ldots a_7}&=&\biggl[i_{\hat{K}}(i_{\hat{\alpha}}\hat{G}^{(9)})+d\hat{\alpha}\wedge i_{\hat{\alpha}}\hat{\Lambda}-d(\hat{R}^2)\wedge \hat{\Lambda}\\ &&i_{\hat{\alpha}}\hat{\omega}\wedge i_{\hat{\alpha}}\hat{F}^{(7)}+i_{\hat{\alpha}}\hat{\Sigma}\wedge i_{\hat{\alpha}}\hat{F}\biggr]_{a_1\ldots a_7}
\end{eqnarray}
where we have multiplied through by $\hat{R}^2$. Combining the $d(\hat{R}^2)$ and $d\hat{\Lambda}$ terms into a single differential and re-expressing everything in the usual co-ordinate basis, which gives us an extra term, we obtain:
\begin{eqnarray}\label{eq:11-d dLambda 3}
\nonumber d(\hat{R}^2\hat{\Lambda})_{\mu_1\ldots \mu_7}&=&\biggl[i_{\hat{K}}(i_{\hat{\alpha}}\hat{G}^{(9)})+d\hat{\alpha}\wedge i_{\hat{\alpha}}\hat{\Lambda}-d(i_{\hat{\alpha}}\hat{\Lambda}))\wedge \hat{\alpha}\\ &&+i_{\hat{\alpha}}\hat{\omega}\wedge i_{\hat{\alpha}}\hat{F}^{(7)}+i_{\hat{\alpha}}\hat{\Sigma}\wedge i_{\hat{\alpha}}\hat{F} \biggr]_{\mu_1\ldots \mu_7}
\end{eqnarray}
Due to the way this equation was constructed at first glance one may expect that it is not valid if a $z$ co-ordinate appears as a free index. In fact, when this does occur it is easy to see that the equation still holds, albeit trivially. Therefore (\ref{eq:11-d dLambda 3}) is fully tensorial. Furthermore, because of its nice structure we can use it to determine the generalised charge for the KK monopole, in contrast to (\ref{eq:11-d dLambda}), i.e. all the terms can be cancelled by taking the exterior derivative of some other set of terms, hence a charge can be constructed.\\
\indent Before we can construct the charge however we must define the field strength equation for $G^{(9)}$. We note that only $i_{\alpha}G^{(9)}$ appears in (\ref{eq:11-d dLambda 3}) and so it is only these components which the potential need be defined for. We then define the potential $i_{\alpha}N^{(8)}$ as:
\begin{eqnarray}\label{eq:11-d N8pot}
\nonumber i_{\hat{\alpha}}\hat{G}^{(9)}&=&-d(i_{\hat{\alpha}}\hat{N}^{(8)})-d(i_{\hat{\alpha}}\hat{A})\wedge i_{\hat{\alpha}}\hat{C}-\frac{1}{6}d(\hat{i_{\alpha}}\hat{A})\wedge i_{\hat{\alpha}}\hat{A}\wedge \hat{A}\\ &&+\frac{1}{6}d\hat{A}\wedge i_{\hat{\alpha}}\hat{A}\wedge i_{\hat{\alpha}}\hat{A}
\end{eqnarray}
in accordance with \cite{Sato:1999bu,Sato:2000mw}, where the gauge transformations of $i_{\hat{\alpha}}\hat{N}^{(8)}$ are also stated. Furthermore, this definition reduces to (\ref{eq:IIA dC7}) upon dimensional reduction along the isometry direction $\hat{\alpha}$. For details of how all the fields reduce refer to Appendix \ref{sec:reducs}. \\
\indent Using the above relations we find the generalised charge for the KK monopole to be
\begin{eqnarray}\label{eq:11-d KK-charge}
\nonumber \hat{L}_{(KK)}&=&\hat{R}^2\hat{\Lambda}+i_{\hat{\alpha}}\hat{\Lambda} \wedge \hat{\alpha}-i_{\hat{K}}(i_{\hat{\alpha}}\hat{N}^{(8)}-\frac{1}{3!}\hat{A}\wedge (i_{\hat{\alpha}}\hat{A})^2)-i_{\hat{\alpha}}\hat{\omega}\wedge (i_{\hat{\alpha}}\hat{C}\\ &&+\frac{1}{2}\hat{A}\wedge i_{\hat{\alpha}}\hat{A})+i_{\hat{\alpha}}\hat{L}_{(5)}\wedge i_{\hat{\alpha}}\hat{A}-\frac{1}{2}\hat{L}_{(2)}\wedge (i_{\hat{\alpha}}\hat{A})^2
\end{eqnarray}
where not only ${\cal L}_{\hat{K}}$ of the potentials is chosen to vanish as usual but it is also assumed that ${\cal L}_{\hat{\alpha}}$ of the potentials vanishes. It is stated in \cite{Eyras:1998kx} that this latter choice is always possible in the IIA case and so it must also  be achievable in the 11-d case.\\
\indent We would expect (\ref{eq:11-d KK-charge}) to have certain characteristics that follow on from the spacetime geometry of the KK monopole solution and interpreting $\hat{\alpha}$ specifically as the transverse Taub-NUT isometry. Firstly, since the isometry lies transverse to the monopole worldvolume, the components parallel to the $z$ direction should vanish. In fact we see that this is indeed the case by noting that $i_{\hat{\alpha}}\hat{L}_{(KK)}=0$. Furthermore, dimensional reduction of the charge specifically along $\hat{\alpha}$ should lead to the D6-brane charge in IIA. This is also found to be the case using the reduction rules given in Appendix \ref{sec:reducs}. Of course, one is also allowed to reduce the KK monopole solution along directions other than the Taub-NUT isometry. One option is to do a double reduction in which case one arrives at the IIA KK monopole, the charge of which is given in the next section. A second option is to perform a direct reduction (but not along $\hat{\alpha}$) in which case one produces an exotic brane called the IIA KK6 monopole (see \cite{Eyras:1999at}). We consider this latter option in the forthcoming paper \cite{Callister:2007}.\\
\indent It is also interesting to note that the charge has an explicit dependence on both $\hat{\alpha}$ and $\hat{R}$. Generally, when isometries are introduced by hand, for example in order to perform the latter two dimensional reductions described in the previous paragraph, the isometry does not appear in the charge since the expression for the charge is valid generally. Here though, $\hat{\alpha}$ is playing the role of a potential and the fact that it appears as part of the charge is not surprising since the Taub-NUT isometry is an essential part of the spacetime solution. The appearance of $\hat{R}^2$ in the charge is also to be expected therefore since this is equal to $|\hat{\alpha}|^2$, but furthermore this factor characteristically appears in the expressions for the KK monopoles' tensions \cite{Bergshoeff:1998re,Bergshoeff:1998bs} and consequently in the worldvolume action multiplying the Born-Infeld term \cite{Bergshoeff:1997gy}. Therefore from our experiences with the 10-d charges its presence in (\ref{eq:11-d KK-charge}) is expected.    

\subsection{KK-monopole charges in 10-dimensions}\label{sec:10-d KK-charges}
The charges for the 10-dimensional KK monopoles can be found in precisely the same manner as for the 11-dimensional case with the minor difference now that the worldvolume is 5-dimensional. This method would essentially involve determining the form of an extra IIA algebraic Killing spinor equation along the lines of (\ref {eq:11-d alg rel}). An algebraic relation could then be obtained from this that, when combined with (\ref{eq:IIA sigmatilde dif}) and (\ref{eq:IIA sigmatilde alg}) would produce a differential relation for $\tilde{\Sigma}$ that's free from any problematic contractions; this would then be the starting point in finding the expression for the KK monopole charge. However, in practice this method becomes rather cumbersome and so we can simply determine the IIA charge by dimensional reduction of (\ref{eq:11-d KK-charge}) along a direction parallel to the worldvolume. Both methods produce the same result which turns out to be:
\begin{eqnarray}\label{eq:IIA KK-charge}
\nonumber M_{(KK)}&=&e^{-2\phi}R^2\tilde{\Sigma}-e^{-2\phi}i_{\alpha}\tilde{\Sigma}\wedge\alpha+e^{-\phi}i_{\alpha}C^{(1)}i_{\alpha}\Lambda\\ \nonumber &&-i_K(i_{\alpha}N^{(7)})+\frac{1}{2}i_K(B\wedge (i_{\alpha}C^{(3)})^2)\\ \nonumber &&+i_{\alpha}(e^{-\phi}\Omega+\tilde{K}\wedge C^{(1)})\wedge(i_{\alpha}C^{(5)}-i_{\alpha}C^{(3)}\wedge B)\\ \nonumber &&+i_{\alpha}\tilde{K}(i_{\alpha}B^{(6)}+\frac{1}{2}C^{(3)}\wedge i_{\alpha}C^{(3)})+i_{\alpha}{\cal M}_{(5)}\wedge i_{\alpha}B\\ &&+i_{\alpha}M_{(4)}\wedge i_{\alpha}C^{(3)}-M_{(2)}\wedge i_{\alpha}C^{(3)}\wedge i_{\alpha}B\nonumber \\ &&-\frac{1}{2}{\cal M}_{(1)}\wedge(i_{\alpha}C^{(3)})^2
\end{eqnarray}
where $N^{(7)}$ is the 7-form potential dual to the IIA Taub-NUT isometry $\alpha$ whose field strength $G^{(8)}=e^{-2\phi}\ast d\alpha$ is found, also by dimensional reduction, to be given by:
\begin{eqnarray}
\nonumber i_{\alpha}G^{(8)}&=&-d(i_{\alpha}N^{(7)})-i_{\alpha}C^{(1)}i_{\alpha}F^{(8)}-i_{\alpha}dC^{(3)}\wedge i_{\alpha}C^{(5)}\\ \nonumber &&-i_{\alpha}H\wedge i_{\alpha}B^{(6)}-\frac{1}{2}i_{\alpha}H\wedge i_{\alpha}C^{(3)}\wedge C^{(3)}\\ &&+\frac{1}{2}H\wedge (i_{\alpha}C^{(3)})^2
\end{eqnarray}
Note that $\hat{R}$, $\hat{\alpha}$ and $i_{\hat{\alpha}}\hat{N}^{(8)}$ have differing reduction rules depending on whether the reduction takes place along $\hat{\alpha}$ or not.  The appropriate rules are given in Appendix \ref{sec:reducs}. In order to show that (\ref{eq:IIA KK-charge}) closes, one needs the differential relation for $\tilde{\Sigma}$, given by
\begin{eqnarray}
\nonumber d(R^2e^{-2\phi}\tilde{\Sigma})&=&d(e^{-2\phi}i_{\alpha}\tilde{\Sigma}\wedge \alpha)+i_K(i_{\alpha}G^{(8)})+e^{-2\phi}i_{\alpha}\Sigma\wedge i_{\alpha}H\\ \nonumber &&-e^{-\phi}\biggl[i_{\alpha}\Lambda\wedge i_{\alpha}F^{(2)}+i_{\alpha}Z\wedge i_{\alpha}F^{(4)}+i_{\alpha}\Omega\wedge i_{\alpha}F^{(6)}\biggr]\\ && +i_{\alpha}\tilde{K}i_{\alpha}H^{(7)}
\end{eqnarray}
which is the double dimensional reduction of (\ref{eq:11-d dLambda 3}).\\
\indent Once again we have chosen gauges so that ${\cal L}_K$ and ${\cal L}_{\alpha}$ of all the potentials vanish, just as in the 11-d theory.\\
\indent Moving on to the IIB case we can repeat the 11-d method. The details of the calculation are long-winded so we do not present the here. Carrying out this process gives the IIB KK monopole charge as 
\begin{eqnarray}\label{eq:IIB KK-charge}
\nonumber N_{(KK)}&=&e^{-2\varphi}{\cal R}^2\Sigma^{+}-e^{-2\varphi}i_{\beta}\Sigma^{+}\wedge \beta-i_K(i_{\beta}{\cal N}^{(7)}) +e^{-\varphi}i_{\beta}\Sigma^{12}\wedge i_{\beta}{\cal C}^{(2)}\\ \nonumber &&+e^{-\varphi}i_{\beta}\Phi^{12}\wedge i_{\beta}{\cal C}^{(4)}+e^{-\varphi}i_{\beta}K^{12}\wedge i_{\beta}{\cal C}^{(6)}-i_{\beta}{\cal N}_{(5)}\wedge i_{\beta}{\cal B}\\ \nonumber &&+i_{\beta}K^-(-i_{\beta}{\cal B}^{(6)}+{\cal C}^{(2)}\wedge i_{\beta}{\cal C}^{(4)}-{\cal C}^{(0)}i_{\beta}{\cal C}^{(6)})\\ \nonumber &&+K^-\wedge i_{\beta}{\cal C}^{(4)}\wedge i_{\beta}{\cal C}^{(2)}+ \frac{1}{2}i_K(i_{\beta}{\cal C}^{(4)})\wedge i_{\beta}{\cal C}^{(4)}\nonumber \\ &&-i_K(i_{\beta}{\cal C}^{(2)})\wedge i_{\beta}{\cal C}^{(6)}
\end{eqnarray}
where ${\cal R}$ is radius of compact direction in IIB. The Taub-NUT isometry here is denoted by the Killing vector $\beta$ and its dual potential ${\cal N}^{(7)}$ has a field strength ${\cal G}^{(8)}=e^{-2\varphi}\ast d\beta$ given by
\begin{eqnarray}\label{eq:IIB G8}
\nonumber i_{\beta}{\cal G}^{(8)}&=&-d(i_{\beta}{\cal N}^{(7)})-i_{\beta}{\cal B}^{(6)}\wedge i_{\beta}{\cal H}-i_{\beta}{\cal C}^{(2)}\wedge i_{\beta}{\cal F}^{(7)}\\ && -\frac{1}{2}i_{\beta}{\cal C}^{(2)}\wedge i_{\beta}{\cal C}^{(4)}\wedge {\cal H} +\frac{1}{2}{\cal C}^{(2)}\wedge i_{\beta}{\cal C}^{(4)}\wedge i_{\beta}{\cal H}\nonumber \\ &&+\frac{1}{2}i_{\beta}{\cal C}^{(4)}\wedge i_{\beta}{\cal F}^{(5)}
\end{eqnarray}
This field strength equation was found by T-dualising (\ref{eq:IIA dB6}). The IIB charge can be shown to be closed by using the following differential relation for $\Sigma^{+}$
\begin{eqnarray}\label{eq:IIB dSigma33}
\nonumber d(e^{-2\varphi}{\cal R}^2\Sigma^{+})&=&d(e^{-2\varphi}i_{\beta}\Sigma^{+}\wedge \beta)+i_{K^{+}}(i_{\beta}{\cal G}^{(8)})-i_{\beta}K^{-}i_{\beta}{\cal H}^{(7)}\\ \nonumber &&+e^{-\varphi}\biggl[ i_{\beta}\Pi^{12} i_{\beta}{\cal F}^{(1)}+i_{\beta}\Sigma^{12}\wedge i_{\beta}{\cal F}^{(3)}+i_{\beta}\Phi^{12}\wedge i_{\beta}{\cal F}^{(5)}\\ && +i_{\beta}K^{12}i_{\beta}{\cal F}^{(7)} \biggr] -e^{-2\varphi}i_{\beta}\Sigma^{-}\wedge i_{\beta}{\cal H}
\end{eqnarray}
which can be found by T-dualising (\ref{eq:IIA dSigma}). Once again, the relevant Lie derivatives are understood to vanish. Note that this means the term $i_{\beta}{\cal F}^{(1)}$ in (\ref{eq:IIB dSigma33}) is equal to ${\cal L}_{\beta}{\cal C}^{(0)}$ and hence zero in this gauge. However, there are instances where this is not zero so we have left this term explicit in (\ref{eq:IIB dSigma33}) for purposes of generality. This situation is briefly discussed in Section \ref{sec:T-dual} where we consider the IIB theory that is the T-dual of the massive IIA theory discussed in the next section.\\
\indent We briefly comment that both of these 10-dimensional KK-charges share the same characteristics as the 11-d case and the discussion there applies here. The IIB monopole tension factor is simply read off as being $R^2e^{-2\phi}$ in agreement with \cite{Eyras:1998hn}. The IIA monopole has a tension $e^{-2\phi}R^2\sqrt{1+e^{2\phi}R^{-2}(i_{\alpha}C^{(1)})^2}$, \cite{Bergshoeff:1998ef}. This can be read of from (\ref{eq:IIA KK-charge}) by treating $i_{\alpha}\Lambda$ as a second 5-form. Note however that this term is bounded by the volume of a 6 dimensional surface with one compact direction with radius $R$. A factor of $R$ therefore has to be introduced when interpreting this form as calibrating  5 dimensional surfaces.

\section{Massive Supergravity}\label{sec:massive supergravity}
We now consider the massive versions of IIA and 11-d supergravity. The ultimate reason for doing this is to find the M9-brane generalised charge which only exists in massive 11-d SUGRA. We consider the IIA theory here also for the sake of completeness and to justify our expression for the D8-brane charge. We begin by describing the massive IIA theory and re-testing the charges already found before repeating the analysis in the 11-d case and then finally finding the M9-charge itself.

\subsection{Massive IIA Supergravity}\label{sec:massive IIA}
It is well known that the IIA supergravity theory we have considered thus far can be consistently modified to include a cosmological constant term $m^2/2$, where $m$ is commonly referred to as the mass parameter. This theory was originally constructed in \cite{Romans:1985tz} by Romans but has also been considered more recently in \cite{Bergshoeff:1996ui} and \cite{Bergshoeff:2006qw}. Here we investigate whether the generalised charges previously constructed for the massless IIA theory, still hold when we consider the massive case.\\
\indent The massive theory, simply put, can be viewed as an extension to the massless version already considered. Essentially the field content is the same except for the presence of $m$ which is naturally interpreted as a scalar Ramond-Ramond field strength and is related to a 10-form field strength $F^{(10)}$ via $m=\ast F^{(10)}$ as noted in \cite{Green:1996bh}. Furthermore the equation of motion for $m$ is found to be $dm=0$, hence $m$ is a constant.\footnote{Actually $m$ only has to be piecewise constant as it can have discontinuities across D8-branes which act as domain walls. See \cite{Bergshoeff:1996ui} for more details.} The definitions of the other fields and the Killing spinor equations already presented now receive additional contributions proportional to $m$. In this section we first state these terms and then differentiate the charges we found in the massless case to see what role they play.\\
\indent  First we consider the Killing spinor equations (\ref{eq:IIA diff rel}) and (\ref{eq:IIA alg rel}) which receive the following additional terms, \cite{Bergshoeff:2006qw}:\footnote{Note that our conventions differ from those in \cite{Bergshoeff:2006qw} by $B\rightarrow-B_{(2)}$, $C^{(1)}\rightarrow-C_{(1)}$, $C^{(5)}\rightarrow-C_{(5)}$ and $C^{(9)}\rightarrow-C_{(9)}$. }
\begin{eqnarray}
\delta \psi _{\mu}&\sim&+\frac{1}{8}e^{\phi}m\Gamma_{\mu}\epsilon \nonumber\\\label{eq:mIIA k spinor}
\delta \lambda &\sim&+\frac{5}{4}e^{\phi}m\epsilon
\end{eqnarray}
which leads to alterations in the differential equations obtained for the bilinears. For the differential relations we found in the massless case we have the following modifications
\begin{eqnarray}\label{eq:mIIA dX alt}
d(e^{-\phi}X)&\sim&-m\tilde{K}\\
d(e^{-2\phi}\Sigma)&\sim&-me^{-\phi}\Lambda\\
d(e^{-\phi}\Psi)&\sim&+m\tilde{\Pi}=-i_KF^{(10)}
\end{eqnarray}
while the others remain unchanged.\\ 
\indent In addition to this the Ramond-Ramond field equations (\ref{eq:IIA dC1}) - (\ref{eq:IIA dC9}) each receive an additional term:
\begin{eqnarray}\label{eq:mIIA RR potentials}
dC^{(2n-1)}&\sim&-\frac{1}{n!}m(B)^n
\end{eqnarray}
with the equation for $H$ remaining unchanged.\\
\indent Now applying these changes to the D0-brane charge (\ref{eq:IIA d0-charge}) as a first example we find\begin{eqnarray}\label{eq:mIIA d0-charge test}
dM_{(0)}&=&-m\tilde{K}-mi_KB-{\cal L}_KC^{(1)}\nonumber \\
&=&-m{\cal M}_{(1)}-{\cal L}_KC^{(1)}
\end{eqnarray}
If we were to pick a gauge where ${\cal L}_KC^{(1)}=0$ as in the massless theory, then $M_{(0)}$ would obviously not be closed. In the massive theory however, such a gauge choice is not generally possible. To see this consider for the moment the massless gauge transformation $C^{(1)}\rightarrow C^{(1)}+d\lambda_{(0)}$. This has the effect of shifting ${\cal L}_KC^{(1)}$ by an exact term but we see from (\ref{eq:mIIA d0-charge test}) that ${\cal L}_KC^{(1)}$ is now no longer exact as it was in the massless theory, so ${\cal L}_KC^{(1)}=0$ cannot be achieved generally. This argument is essentially the same as we presented in Section \ref{sec:11-d charges} for the gauge choice of $\hat{A}$, where there it was used to show ${\cal L}_{\hat{K}}\hat{A}=0$ was viable. If we perform the analogous transformation here we obtain the condition
\begin{eqnarray}\label{eq:mIIA C1 gauge}
{\cal L}_KC^{(1)}=-m{\cal M}_{(1)}
\end{eqnarray}
for which $M_{(0)}$ does close. This condition can be thought of as a generalisation of the massless case for non-zero $m$.\footnote{Note that now there is also a massive gauge transformation that can be performed on $C^{(1)}$, which gives slightly more freedom to the possible gauge choices available. However, these involve the gauge parameter associated with gauge transformations of $B$ and so are somewhat diminished if we fix a gauge condition for $B$. (See \cite{Bergshoeff:2006qw} for a full list of the gauge transformations.) In any case, here we are not really concerned with what gauge choices are not possible for $C^{(1)}$, but rather that a gauge choice is possible where $M_{(0)}$ is closed, namely (\ref{eq:mIIA C1 gauge})}. This gauge choice has a knock-on effect for the higher rank potentials. For consistency we have
\begin{eqnarray}
{\cal L}_KC^{(2n-1)}=-\frac{1}{(n-1)!}m{\cal M}_{(1)}\wedge (B)^{(n-1)}
\end{eqnarray}
With these gauge conditions all the D-brane charges presented previously are found to be closed in the massive version of the theory. Note that in the case of the D8-brane the field strength $F^{(10)}$ appears both in the derivatives of the potential $C^{(9)}$ and the 8-form bilinear $\Psi$, and both appearances cancel in the D8-brane charge.\\
\indent Looking next at the NS branes, we find for the F-string that the charge is still closed for ${\cal L}_KB=0$ since neither the exterior derivative of $B$ or $\tilde{K}$ receive massive corrections. We cannot comment too much on the NS9-brane case since we are not completely sure of the expression for the charge but it is reasonable to assume it is unchanged from the massless version as well, since ${\cal L}_KD^{(10)}$ would be exact and hence could be chosen to vanish.\\
\indent Before considering the NS5-brane we state the massive extension to the field equation for $B^{(6)}$ which from \cite{Bergshoeff:1997ak,Bergshoeff:2006qw} is:\footnote{Again our definition of $B^{(6)}$ differs from that in \cite{Bergshoeff:2006qw} by $B^{(6)}\rightarrow B_{(6)}+C_{(5)}\wedge C_{(1)}-\frac{1}{2}C_{(3)}\wedge C_{(1)}\wedge B$, and also from \cite{Bergshoeff:1997ak} by $m\rightarrow m/2$.}
\begin{eqnarray}\label{eq:mIIA dB6}
\nonumber dB^{(6)}&=&H^{(7)}-C^{(1)}\wedge F^{(6)}+\frac{1}{2}C^{(3)}\wedge (F^{(4)}+H\wedge C^{(1)})\\ &&+m(C^{(7)}-C^{(5)}\wedge B+\frac{1}{4}C^{(3)}\wedge(B)^2)
\end{eqnarray} 
Then we find that (\ref{eq:IIA NS5-charge}) is still closed if the gauge choice for $B^{(6)}$ is made so that
\begin{eqnarray}
{\cal L}_KB^{(6)}&=&-m{\cal M}_{(1)} \wedge C^{(5)}+\frac{1}{2}m{\cal M}_{(1)} \wedge C^{(3)}\wedge B+mM_{(6)}
\end{eqnarray}
In this instance the gauge transformations are more complicated than for the RR potentials, but we can still shift ${\cal L}_KB^{(6)}$ by an exact term, hence the essential point of the argument given previously applies here to show that this gauge condition is possible.\\ 
\indent Finally we turn our attention to the KK-monopole. In this case we need to know how the field strength equation for $i_{\alpha}N^{(7)}$ is modified for non-zero $m$. The authors have been unable to find this field strength equation for the massive IIA theory in the literature explicitly, though it was indirectly stated in \cite{Eyras:1998kx} through the massive gauge transformations of $i_{\alpha}N^{(7)}$. With our definitions it should take the following form: 
\begin{eqnarray}\label{eq:mIIA dN7}
i_{\alpha}G^{(8)}&=&-d(i_{\alpha}N^{(7)})-i_{\alpha}C^{(1)}i_{\alpha}F^{(8)}-i_{\alpha}H\wedge i_{\alpha}B^{(6)}\nonumber \\ &&-(i_{\alpha}F^{(4)}+i_{\alpha}H\wedge C^{(1)}-i_{\alpha}C^{(1)}H)\wedge i_{\alpha}C^{(5)}\nonumber \\ &&-\frac{1}{2}i_{\alpha}H\wedge i_{\alpha}C^{(3)}\wedge C^{(3)}+\frac{1}{2}H\wedge (i_{\alpha}C^{(3)})^2\nonumber \\ &&+m(i_{\alpha}B\wedge i_{\alpha}C^{(7)}+\frac{1}{2}i_{\alpha}B\wedge (B)^2\wedge i_{\alpha}C^{(3)})
\end{eqnarray}
Differentiating this yields
\begin{eqnarray}
d(i_{\alpha}G^{(8)})&=&i_{\alpha}F^{(2)}\wedge i_{\alpha}F^{(8)}-i_{\alpha}F^{(4)}\wedge i_{\alpha}F^{(6)}+i_{\alpha}H\wedge i_{\alpha}H^{(7)}
\end{eqnarray}
which seems like a feasible modified Bianchi identity for $i_{\alpha}G^{(8)}$, most notably because it is manifestly gauge invariant. Using this relation we find that (\ref{eq:IIA KK-charge}) is closed if we make the following gauge choice for $i_{\alpha}N^{(7)}$
\begin{eqnarray}
{\cal L}_K(i_{\alpha}N^{(7)})&=&-mi_{\alpha}{\cal M}_{(1)}\wedge i_{\alpha}C^{(7)}+mi_{\alpha}B\wedge i_{\alpha}M_{(6)}\nonumber \\ &&-m\frac{1}{2}i_{\alpha}({\cal M}_{(1)} \wedge (B)^2)\wedge i_{\alpha}C^{(3)}
\end{eqnarray} 
We therefore see that all the massless charges are still valid in the massive theory provided the appropriate gauge conditions are met, which although more complicated, reduce to the massless conditions for $m=0$.

\subsection{Massive 11 dimensional Supergravity}\label{sec:massive 11 dimensional supergravity}
We will now consider the massive 11-d theory presented in \cite{Bergshoeff:1997ak}. As in the IIA case, for our purposes we can simply consider the massive version here to be a straight forward extension of the massless theory already considered but with an additional scalar field $\hat{m}$. One additional complication however is that for consistency there is also a Killing isometry present in the theory. The necessity of this isometry arises due to the no-go theorem presented in \cite{Bautier:1997yp} which prevents the formulation of a generally covariant massive 11-d theory. The presence of the isometry allows the introduction of a mass parameter by sacrificing covariance and is therefore an integral part of the theory. It makes explicit appearances in the extra massive terms in the Killing spinor and field equations.\\
\indent The source of the mass parameter is the M9-brane and as such it can be related to an 11-form field strength, $|\hat{\alpha}|^4\hat{m}=\hat{\ast} \hat{F}^{(11)}$, where $\hat{\alpha}$ is the `massive' isometry. Dimensional reduction along the massive isometry leads to the massive IIA theory considered in the last section. A reduction can be performed along a different direction to this but then one arrives at a non-covariant IIA supergravity which will not be considered in this paper. Note that by setting $\hat{m}=0$ and restoring dependence along the isometry, one recovers the standard massless theory.\\
\indent We now consider how the massless equations are changed for the massive theory. The extra massive terms in the Killing spinor equation (\ref{eq:11-d killingspinor}) should reduce to (\ref{eq:mIIA k spinor}) and are found to be
\begin{eqnarray}\label{eq:m11-d diff rel}
\hat{\tilde{D}}_a\epsilon &\sim& +\hat{m}\hat{R}^2(\frac{5}{12}\hat{\Gamma}_a-\frac{1}{2}\hat{\Gamma}_{\underline{z}}\hat{\Gamma}_{\underline{z}a})\hat{\epsilon}+\frac{1}{4}\hat{m}\hat{R}i_{\hat{\alpha}}\hat{A}_{ab}\hat{\Gamma}^b_{\phantom{N}\underline{z}}\hat{\epsilon} \nonumber \\
 && +\frac{1}{8}\hat{m}\hat{R}i_{\hat{\alpha}}\hat{A}_{b_1b_2}(\hat{\Gamma}^{b_1b_2}_{\phantom{NP}a\underline{z}}-\hat{\Gamma}^{b_1b_2}_{\phantom{b_1b_2}a}\hat{\Gamma}_{\underline{z}})\hat{\epsilon}
\end{eqnarray}
where we work in the orthonormal basis adapted to the isometry $z$ with $\hat{\alpha}^{\mu}=\delta^{\mu z}$ and $\hat{R}$ is the radius of the isometry direction which we assume to be compact. Note that the Roman indices here run over all values. Effectively the above expression acts like a conditional  statement, giving two different sets of terms depending on whether $a=\underline{z}$ or not. This seems to be the only way to `unify' the massive IIA terms (\ref{eq:mIIA k spinor}). Consequently the 11-d algebraic Killing spinor equation (\ref{eq:11-d alg rel}) is modified to
\begin{eqnarray}\label{eq:m11-d alg rel}
\hat{\tilde{D}}_z\hat{\epsilon} \sim +\frac{5}{12}\hat{m}\hat{R}^3\hat{\Gamma}_{\underline{z}}\hat{\epsilon}-\frac{1}{8}\hat{m}\hat{R}^2i_{\hat{\alpha}}\hat{A}_{b_1b_2}\hat{\Gamma}^{b_1b_2}\hat{\epsilon}
\end{eqnarray}
Here we have assumed that the isometry used to construct the algebraic Killing spinor relation is the massive isometry $\hat{\alpha}$. Of course, if another isometry were used then the additional massive terms could still be found from (\ref{eq:m11-d diff rel}), but they would not simplify to (\ref{eq:m11-d alg rel}). We do not consider that option in this paper, but see \cite{Callister:2007}.\\
\indent Note the appearance of the 3-form potential in the above expressions. This seems strange since the Killing spinor equations are now not gauge invariant. This is due to the general property of this theory in that objects that would normally be gauge invariant in the massless case, now become gauge covariant under massive transformations. The 3-form potential above arises because the connection in the covariant derivative in (\ref{eq:11-d killingspinor}) now needs to be extended for the massive gauge transformations. We will not worry about the details of this but see \cite{Bergshoeff:1997ak} for more information.\\
\indent With these modifications we find the differential relations for the bilinear forms are also modified. The changes we need to consider here are summarised as 
\begin{eqnarray}
d\hat{\omega}&\sim&-\hat{m}i_{\hat{\alpha}}\hat{A}\wedge i_{\hat{\alpha}}\hat{\omega}\\
d\hat{\Sigma}&\sim&-\hat{m}(\hat{R}^2\hat{\Lambda}+i_{\hat{\alpha}}\hat{\Lambda}\wedge \hat{\alpha}+i_{\hat{\alpha}}\hat{A}\wedge i_{\hat{\alpha}}\hat{\Sigma})\\
d(\hat{R}^2\hat{\Lambda}+i_{\hat{\alpha}}\hat{\Lambda}\wedge \hat{\alpha})&\sim&0\\\label{eq:m11-d dPi}
i_{\hat{\alpha}}d\hat{\Pi}&\sim&-\frac{2}{3}\hat{R}^{-2}i_{\hat{K}}(i_{\hat{\alpha}}\hat{F}^{(11)})
\end{eqnarray}
Also note that now $\hat{K}$ is no longer Killing because $\hat{\nabla}_{(\mu}\hat{K}_{z)}\neq0$.\\
\indent Extra terms are also present in the field equations and are given by \cite{Bergshoeff:1997ak,Sato:1999bu}:
\begin{eqnarray}
\hat{G}^{(2)}&=&d\hat{\alpha}+\hat{m}\hat{R}^2i_{\hat{\alpha}}\hat{A}\\
\hat{F}^{(4)}&=&d\hat{A}+\frac{1}{2}\hat{m}(i_{\hat{\alpha}}\hat{A})^2\\
\hat{F}^{(7)}&=&d\hat{C}-\frac{1}{2}d\hat{A}\wedge \hat{A}+\hat{m}i_{\hat{\alpha}}\hat{A}\wedge i_{\hat{\alpha}}\hat{C}-\frac{1}{3!}\hat{m}\hat{A}\wedge (i_{\hat{\alpha}}\hat{A})^2\nonumber \\ &&+\hat{m}i_{\hat{\alpha}}\hat{N}^{(8)}\\\label{eq:m11-d N8 pot}
i_{\hat{\alpha}}\hat{G}^{(9)}&=&-d(i_{\hat{\alpha}}\hat{N}^{(8)})+i_{\hat{\alpha}}(d\hat{A})\wedge i_{\hat{\alpha}}\hat{C}+\frac{1}{6}i_{\hat{\alpha}}(d\hat{A}\wedge i_{\hat{\alpha}}\hat{A}\wedge \hat{A})\nonumber \\ &&+\frac{1}{4!}\hat{m}(i_{\hat{\alpha}}\hat{A})^4
\end{eqnarray}
These can be seen to reduce down to give the correct field relations in IIA. It should be stressed that $\hat{\alpha}$ in the equations above is specifically the massive isometry, and hence the reduction must take place along this direction in order to arrive at massive IIA supergravity. Furthermore, here $\hat{N}^{(8)}$ is specifically dual to $\hat{\alpha}$ and as such does not dimensionally reduce to $N^{(7)}$ in IIA. It is possible to consider the field equation for an 8-form potential dual to a general isometry, i.e. not $\hat{\alpha}$, which will be partly the same as (\ref{eq:m11-d N8 pot}) but with differing massive terms. We consider such a field in the forthcoming article \cite{Callister:2007}.\\
\indent One also has a non-dynamical 10-form potential $\hat{A}^{(10)}$ for the 11-form field strength $\hat{F}^{(11)}$ which can be introduced into the action as an auxiliary field. See \cite{Sato:1999bu,Bergshoeff:1996ui}.  We define the field equation as:
\begin{eqnarray}\label{eq:m11-d A10 pot}
\nonumber i_{\hat{\alpha}}\hat{F}^{(11)}&=&-d(i_{\hat{\alpha}}\hat{A}^{(10)})-i_{\hat{\alpha}}\hat{F}\wedge i_{\hat{\alpha}}\hat{N}^{(8)}+\frac{1}{4!}i_{\hat{\alpha}}(d\hat{A}\wedge \hat{A}\wedge (i_{\hat{\alpha}}\hat{A})^2)\\ && +\frac{1}{5!}\hat{m}(i_{\hat{\alpha}}\hat{A})^5
\end{eqnarray}
This equation was constructed so that it reduces to the field equation for $C^{(9)}$ in IIA using the rules given in Appendix \ref{sec:reducs}.\\
\indent Using these new relations we can check to see whether our charges for the massless 11-d case are still closed. We find a similar situation to that encountered for the massive IIA theory, where the charges do remain closed if the appropriate gauge conditions are chosen. These are given by
\begin{eqnarray}
{\cal L}_{\hat{K}}\hat{A}&=&\hat{m}i_{\hat{\alpha}}\hat{A}\wedge i_{\hat{\alpha}}\hat{L}_{(2)}\\
{\cal L}_{\hat{K}}C&=&-\hat{m}i_{\hat{\alpha}}\hat{C}\wedge i_{\hat{\alpha}}\hat{L}_{(2)}+\hat{m}\hat{L}_{(KK)}\\
{\cal L}_{\hat{K}}(i_{\hat{\alpha}}\hat{N}^{(8)})&=&0
\end{eqnarray}
which generalise the massless case. It can be shown that such conditions are valid using the argument given in the IIA case. We have also used the fact that ${\cal L}_{\hat{\alpha}}$ of the potentials is always zero, since this is an assumption of the supergravity, \cite{Bergshoeff:1997ak}. Note that for the KK monopole there are two scenarios depending on whether the Taub-NUT and massive isometries are thought to coincide or not. Due to our construction of the algebraic Killing spinor equation (\ref{eq:m11-d alg rel}) and the fact that we have only considered the potential dual to $\hat{\alpha}$, in this paper we only consider the case where the massive isometry and Taub-NUT isometry coincide. The other case will be considered in \cite{Callister:2007}.

\subsection{M9-brane charge}\label{sec:M9-charge}
We now determine the structure of the M9-brane generalised charge. This brane solution was constructed in \cite{Bergshoeff:1998bs} and it only exists in this massive version of 11-d supergravity. It wraps a compact isometry which appears gauged in the worldvolume effective action. As with the KK monopole solution, here one can consider whether or not the wrapped isometry coincides with the massive isometry or not. Once again we only consider the case where the two isometries do coincide but the other case will be discussed in \cite{Callister:2007}. Reducing the M9-brane along $\hat{\alpha}$ produces the IIA D8-brane and so this should be reflected in the M9-charge.\\
\indent The method to construct its charge follows the same lines as that used to construct the KK monopole charges. Since there is always an isometry present we can use the algebraic Killing spinor equation (\ref{eq:11-d alg rel}) with the massive modification (\ref{eq:m11-d alg rel}). This time we note however that the M9-brane wraps the isometry. We once again work in a co-ordinate system where $\hat{\alpha}^{\mu}=\delta^{\mu x}$ and begin by hitting (\ref{eq:11-d alg rel}) (modified by (\ref{eq:m11-d alg rel})) from the left with $\hat{\Gamma}_{a_1\ldots a_9}$, $a_i\neq \underline{x}$, which, after some slight manipulation, yields:
\begin{eqnarray}
\nonumber 0&=&\biggl[-\frac{1}{3}i_{\hat{\alpha}}\hat{\Lambda}\wedge \hat{F}+\frac{2}{3}\hat{\Lambda}\wedge i_{\hat{\alpha}}\hat{F}+\hat{R}^{-2}i_{\hat{\alpha}}\hat{\omega}\wedge i_{\hat{\alpha}}\hat{G}^{(9)}-\hat{R}^{-2}d(\hat{R}^{2})\wedge i_{\hat{\alpha}}\hat{\Pi}\nonumber \\ &&-\frac{5}{3}\hat{R}^{-2}i_{\hat{K}}(i_{\hat{\alpha}}\hat{F}^{(11)})\biggr]_{a_1\ldots a_9}
\end{eqnarray}
Combining this with (\ref{eq:11-d dPi}) contracted with $\hat{\alpha}$ and taking into account (\ref{eq:m11-d dPi}) one arrives at:
\begin{eqnarray}\label{eq:m11-d dPi 2}
d(\hat{R}^2i_{\hat{\alpha}}\hat{\Pi})_{a_1\ldots a_9}=\biggl[\hat{R}^2i_{\hat{\alpha}}\hat{F}\wedge \hat{\Lambda}+i_{\hat{\alpha}}\hat{\omega}\wedge i_{\hat{\alpha}}\hat{G}^{(9)}-i_{\hat{K}}(i_{\hat{\alpha}}\hat{F}^{(11)})\biggr]_{a_1\ldots a_9}
\end{eqnarray}
Rewriting this in the coordinate basis gives us an extra term and yields a useful differential relation:
\begin{eqnarray}
d(\hat{R}^2i_{\hat{\alpha}}\hat{\Pi})&=&\hat{R}^2i_{\hat{\alpha}}\hat{F}\wedge \hat{\Lambda}+i_{\hat{\alpha}}\hat{F}\wedge i_{\hat{\alpha}}\hat{\Lambda}\wedge \hat{\alpha}+i_{\hat{\alpha}}\hat{\omega}\wedge i_{\hat{\alpha}}\hat{G}^{(9)}\nonumber \\ &&-i_{\hat{K}}(i_{\hat{\alpha}}\hat{F}^{(11)})
\end{eqnarray}
The M9-brane charge then turns out to be:
\begin{eqnarray}\label{eq:M9-charge}
\nonumber i_{\hat{\alpha}}\hat{L}_{(9)}&=&\hat{R}^2i_{\hat{\alpha}}\hat{\Pi}+i_{\hat{K}}(i_{\hat{\alpha}}\hat{A}^{(10)}-\frac{1}{4!}\hat{A}\wedge (i_{\hat{\alpha}}\hat{A})^3)\\ \nonumber  &&-i_{\hat{\alpha}}\hat{\omega}\wedge (i_{\hat{\alpha}}\hat{N}^{(8)}-\frac{1}{3!}\hat{A}\wedge (i_{\hat{\alpha}}\hat{A})^2)+\hat{L}_{(KK)}\wedge i_{\hat{\alpha}}\hat{A}\\ &&-\frac{1}{2}i_{\hat{\alpha}}\hat{L}_{(5)}\wedge (i_{\hat{\alpha}}\hat{A})^2+\frac{1}{3!}\hat{L}_{(2)}\wedge (i_{\hat{\alpha}}\hat{A})^3
\end{eqnarray}
where the previous gauge conditions apply along with ${\cal L}_{\hat{K}}(i_{\hat{\alpha}}\hat{A}^{(10)})=0$.\\
\indent From the structure of (\ref{eq:M9-charge}) it is fairly straightforward to see it reduces to give the D8-charge. Note that in order to reduce the brane along a direction different to the isometry that it wraps and still produce Roman's massive IIA theory, one would need to make the massive isometry distinct from the wrapped isometry. Doing this and reducing either transverse or parallel to the worldvolume would produce either a wrapped NS9-brane or the KK8 monopole in IIA respectively, \cite{Bergshoeff:1998bs,Sato:2000mw,Eyras:1999at}. Since we assumed in our construction of (\ref{eq:M9-charge}) that the two isometries coincide we cannot reduce it in these ways here if we want to make contact with the Romans massive IIA supergravity. Note also that although we called the M9-charge $i_{\hat{\alpha}}\hat{L}_{(9)}$, since it is effectively an 8-form due to the isometry it wraps, we have not actually shown that there is a closed 9-form $\hat{L}_{(9)}$ in general. These subjects will be discussed in \cite{Callister:2007}. A final comment on (\ref{eq:M9-charge}) is that it contains the factor $\hat{R}^2$ whereas the brane tension contains the factor $\hat{R}^3$, \cite{Bergshoeff:1998bs,Bergshoeff:1998re}. This situation is similar to the IIA KK monopole case where the extra factor of $R$ arises since $i_{\hat{\alpha}}\hat{\Pi}$ is being pulled back to a surface with a compact direction with radius $R$.     

\section{T-duality of the charges}\label{sec:T-dual}

We now conclude our analysis of the generalised charges by looking at their T-duality relations. We mainly consider the massless IIA theory in this section but comment on the extension to the massive IIA supergravity.

\subsection{Massless case}
It is well known that if the IIA theory has an $S^1$ isometry then one can
perform a T-duality transformation along this isometry to arrive at
the IIB theory and vice-versa. Mathematically speaking, this process is effectively the same
as performing a Kaluza-Klein reduction, in fact the exact nature of
the mapping can be found by inequivalently dimensionally reducing
both theories to the same (unique) maximally supersymmetric 9-dimensional
supergravity theory. This was done explicitly in \cite{Bergshoeff:1995as} and the
(massless) T-duality rules for the lower rank potentials were found.
Subsequently the transformation rules for the higher rank
potentials were also given in \cite{Eyras:1998hn,Eyras:1998kx}. Here we restate these results in our conventions to give the complete set of rules necessary for our purposes.\\
\indent Working in co-ordinate systems adapted to the isometry, in the IIA theory we denote the metric as $g$ , the Killing vector as $\alpha$ and the co-ordinate along the isometry $x$; whilst for the IIB case we have $j$, $\beta$ and $y$ respectively. Also note that in this section the co-ordinates $\mu_i$ do not include the isometry co-ordinates, since we have to separate those co-ordinates. The T-duality rules for the metric going from IIA to IIB then are
\begin{eqnarray}
\nonumber g_{\mu\nu}&\rightarrow&j_{\mu \nu}-(j_{\mu y}j_{\nu y}-{\cal B}_{\mu y}{\cal B}_{\nu y})/j_{yy}\\
\nonumber g_{\mu x}&\rightarrow&-{\cal B}_{\mu y}/j_{yy}\\
g_{xx}&\rightarrow&j_{yy}^{-1}
\end{eqnarray}
and similarly from IIB to IIA:
\begin{eqnarray}
\nonumber j_{\mu\nu}&\rightarrow&g_{\mu \nu}-(g_{\mu x}g_{\nu x}-B_{\mu x} B_{\nu x})/g_{xx}\\
\nonumber j_{\mu y}&\rightarrow&-B_{\mu x}/g_{xx}\\
j_{yy}&\rightarrow&g_{xx}^{-1}
\end{eqnarray}
which obviously hold in these special co-ordinates only.

The T-duality for the fields going from IIA to IIB are
\begin{eqnarray}
\nonumber \phi&\rightarrow&\varphi-\frac{1}{2}\log ({\cal R}^2)\\
\nonumber i_{\alpha}B_{\mu}&\rightarrow&{\cal R}^{-2} \beta_{\mu}\\
\nonumber B_{\mu_1\mu_2}&\rightarrow&({\cal B}+{\cal R}^{-2}i_{\beta}{\cal B}\wedge \beta)_{\mu_1\mu_2}
\\
\nonumber i_{\alpha}C^{(2n+1)}_{\mu_1\ldots \mu_{2n}}&\rightarrow&({\cal C}^{(2n)}+{\cal R}^{-2}i_{\beta}{\cal C}^{(2n)}\wedge \beta)_
{\mu_1\ldots \mu_{2n}}\\
\nonumber
C^{(2n+1)}_{\mu_1\ldots \mu_{2n+1}}&\rightarrow&(-i_{\beta}{\cal C}^{(2n+2)}+{\cal C}^{(2n)}\wedge i_{\beta}{\cal
B}-{\cal R}^{-2}i_{\beta}{\cal C}^{(2n)}\wedge i_{\beta}{\cal B}\wedge \beta)_{\mu_1\ldots \mu_{2n+1}}\\
\nonumber i_{\alpha}B^{(6)}_{\mu_1 \ldots \mu_5}&\rightarrow&(i_{\beta}{\cal B}^{(6)}-\frac{1}{2}i_{\beta}{\cal C}^{(4)}\wedge {\cal C}^{(2)}-\frac{1}{2}{\cal R}^{-2}i_{\beta}{\cal C}^{(4)}\wedge i_{\beta}{\cal C}^{(2)}\wedge \beta)_{\mu_1\ldots \mu_5}\\
\nonumber B^{(6)}_{\mu_1\ldots \mu_6}&\rightarrow&(-i_{\beta}{\cal N}^{(7)}-i_{\beta}{\cal B}^{(6)}\wedge i_{\beta} {\cal B}+\frac{1}{2}i_{\beta}{\cal C}^{(4)}\wedge {\cal C}^{(2)}\wedge i_{\beta}{\cal B}\\ &&-\frac{1}{2}{\cal R}^{-2}i_{\beta}{\cal C}^{(4)}\wedge i_{\beta}{\cal C}^{(2)}\wedge i_{\beta}{\cal B}\wedge \beta)_{\mu_1\ldots \mu_6}\nonumber \\
\nonumber i_{\alpha}N^{(7)}_{\mu_1\ldots \mu_6}&\rightarrow&(-{\cal B}^{(6)}-{\cal R}^{-2}i_{\beta}{\cal B}^{(6)}\wedge \beta+{\cal C}^{(4)}\wedge {\cal C}^{(2)}\\ &&+{\cal R}^{-2}i_{\beta}{\cal C}^{(4)}\wedge {\cal C}^{(2)}\wedge \beta +{\cal R}^{-2}{\cal C}^{(4)}\wedge i_{\beta}{\cal C}^{(2)}\wedge \beta)_{\mu_1\ldots \mu_6}
\end{eqnarray}
whilst going from IIB to IIA we have
\begin{eqnarray}
\nonumber \varphi&\rightarrow&\phi-\frac{1}{2}\log (R^2)\\
\nonumber i_{\beta}{\cal B}_{\mu}&\rightarrow&R^{-2}\alpha_{\mu}\\
\nonumber {\cal B}_{\mu_1\mu_2}&\rightarrow&(B+R^{-2}i_{\alpha}B\wedge \alpha)_{\mu_1\mu_2}\\
\nonumber i_{\beta}{\cal C}^{(2n)}_{\mu_1\ldots \mu_{2n-1}}&\rightarrow&(-C^{(2n-1)}+R^{-2}i_{\alpha}C^{(2n-1)}\wedge \alpha)_{\mu_1\ldots \mu_{2n-1}}\\
\nonumber {\cal C}^{(2n)}_{\mu_1\ldots \mu_{2n}}&\rightarrow&(i_{\alpha}C^{(2n+1)}+C^{(2n-1)}\wedge i_{\alpha}B+R^{-2}i_{\alpha}C^{(2n-1)}\wedge i_{\alpha}B\wedge \alpha)_{\mu_1\ldots \mu_{2n}}\\
\nonumber i_{\beta}{\cal B}^{(6)}_{\mu_1\ldots \mu_5}&\rightarrow&(i_{\alpha}B^{(6)}-\frac{1}{2}i_{\alpha}C^{(3)}\wedge C^{(3)}+\frac{1}{2}R^{-2}i_{\alpha}C^{(3)}\wedge i_{\alpha}C^{(3)}\wedge \alpha)_{\mu_1\dots \mu_5}\\
\nonumber {\cal B}^{(6)}_{\mu_1\ldots \mu_6}&\rightarrow&(-i_{\alpha}N^{(7)}-i_{\alpha}B^{(6)}\wedge i_{\alpha}B +i_{\alpha}C^{(3)}\wedge i_{\alpha}C^{(5)}\nonumber \\ &&+\frac{1}{2}i_{\alpha}C^{(3)}\wedge C^{(3)}\wedge i_{\alpha}B+\frac{1}{2}R^{-2}i_{\alpha}C^{(3)}\wedge i_{\alpha}C^{(3)}\wedge i_{\alpha}B\wedge \alpha)_{\mu_1\ldots \mu_6}\nonumber \\
i_{\beta}{\cal N}^{(7)}_{\mu_1\ldots \mu_6}&\rightarrow&(-B^{(6)}-R^{-2}i_{\alpha}B^{(6)}\wedge \alpha)_{\mu_1\ldots \mu_6}
\end{eqnarray}
where $N^{(7)}$ and ${\cal N}^{(7)}$ are the potentials dual to $\alpha$ and $\beta$ respectively, the isometries over which the T-duality is being performed.\\
\indent The spinors transform, up to normalisation, as
\begin{eqnarray}
\epsilon^+\rightarrow\frac{1}{\sqrt{2}}\epsilon^2 \qquad
\epsilon^-\rightarrow\frac{1}{\sqrt{2}}\Gamma_{\underline{y}}\epsilon^1
\end{eqnarray}
where we have chosen our normalisation for convenience in the T-duality rules for the bilinears given below. Note that $\Gamma_x\rightarrow{\cal R}^{-2}\Gamma_y$. This gives the following transformation rules for the bilinears from IIA to IIB
\begin{eqnarray}
\nonumber K_{\mu}&\rightarrow&K^{+}_{\mu}-{\cal R}^{-2}i_{\beta}K^{+}\beta _{\mu}-{\cal R}^{-2}i_{\beta}K^{-}i_{\beta}{\cal B}_{\mu}\\
\nonumber i_{\alpha}K&\rightarrow&-{\cal R}^{-2}i_{\beta}K^{-}\\
\nonumber\tilde{K}_{\mu}&\rightarrow&-K^{-}_{\mu}+{\cal R}^{-2}i_{\beta}K^{-}\beta _{\mu}+{\cal R}^{-2}i_{\beta}K^{+}i_{\beta}{\cal B}_{\mu}\\
\nonumber i_{\alpha}\tilde{K}&\rightarrow&{\cal R}^{-2}i_{\beta}K^{+}\\
\nonumber Z_{\mu_1\ldots \mu_4}&\rightarrow&{\cal R}^{-1}(-i_{\beta}\Sigma^{12}+\Phi^{12}\wedge i_{\beta}{\cal B}+{\cal R}^{-2}i_{\beta}\Phi^{12}\wedge i_{\beta}{\cal B}\wedge \beta)_{\mu_1\ldots \mu_4}\\
i_{\alpha}Z_{\mu_1\mu_2\mu_3}&\rightarrow&{\cal R}^{-1}(-\Phi^{12}+{\cal R}^{-2}i_{\beta}\Phi^{12}\wedge \beta)_{\mu_1\mu_2\mu_3} 
\end{eqnarray}
and also from IIB to IIA
\begin{eqnarray}
\nonumber K^{+}_{\mu}&\rightarrow&K_{\mu}-R^{-2}i_{\alpha}K\alpha _{\mu}+R^{-2}i_{\alpha}\tilde{K}i_{\alpha}B_{\mu}\\
\nonumber i_{\beta}K^{+}&\rightarrow&R^{-2}i_{\alpha}\tilde{K}\\
\nonumber K^{-}_{\mu}&\rightarrow&-\tilde{K}_{\mu}+R^{-2}i_{\alpha}\tilde{K}\alpha _{\mu}-R^{-2}i_{\alpha}Ki_{\alpha}B_{\mu}\\
\nonumber i_{\beta}K^{-}&\rightarrow&-R^{-2}i_{\alpha}K\\
\nonumber \Sigma^{12}_{\mu_1\ldots \mu_5}&\rightarrow&R^{-1}(-i_{\alpha}\Lambda-Z\wedge i_{\alpha}B+ R^{-2}i_{\alpha}Z\wedge i_{\alpha}B\wedge \alpha)_{\mu_1\ldots \mu_4}\\
i_{\beta}\Sigma^{12}_{\mu_1\ldots \mu_4}&\rightarrow& R^{-1}(-Z- R^{-2}i_{\alpha}Z\wedge \alpha)_{\mu_1\ldots \mu_4} 
\end{eqnarray}
The transformation rules for the other bilinears can be read from the appropriate line here by considering the bilinear with the same number of Dirac matrices mod 2. So for example the transformation rule of $\Sigma$ in IIA is the same as for $K$ but with $K^+\rightarrow \Sigma^+$ and $K^-\rightarrow \Sigma^-$.

\indent The T-duality transformations of the branes have been discussed in
the literature many times. The general rule is that T-dualising a D$n$-brane along it's worldvolume will yield a D$(n-1)$-brane in the dual theory. Conversely, performing the the T-duality transverse to the worldvolume transforms a D$n$-brane into a D$(n+1)$-brane in the dual theory.\footnote{We do not concern ourselves with the technicalities requiring that branes are smeared along isometries transverse to their worldvolumes.} We can T-dualise our D-brane charges to see if they obey the same relationships. When doing this the following relations derived from the equations above prove useful
\begin{eqnarray}
\nonumber i_KC^{(2n+1)}_{\mu_1\ldots \mu_{2n}}&\rightarrow&\biggl[-i_{K^{+}}(i_{\beta}{\cal C}^{(2n+2)})-i_{\beta}K^{-}({\cal C}^{(2n)}+{\cal R}^{-2}i_{\beta}{\cal C}^{(2n)}\wedge \beta)\\ \nonumber && +i_{K^{+}}({\cal C}^{(2n)}+{\cal R}^{-2}i_{\beta}{\cal C}^{(2n)}\wedge \beta)\wedge i_{\beta}B\biggr]_{\mu_1\ldots \mu_{2n}}\\
i_{\alpha}(i_KC^{(2n+1)})_{\mu_1\ldots \mu_{2n-1}}&\rightarrow&-i_{K^{+}}({\cal C}^{(2n)}+{\cal R}^{-2}i_{\beta}{\cal C}^{(2n)}\wedge \beta)_{\mu_1\ldots \mu_{2n-1}}
\end{eqnarray}
\begin{eqnarray}
\nonumber i_{K^{+}}{\cal C}^{(2n)}_{\mu_1\ldots \mu_{2n-1}}&\rightarrow&\biggl[i_K(i_{\alpha}C^{(2n+1)})+i_{\alpha}\tilde{K}(-C^{(2n-1)}+R^{-2}i_{\alpha}C^{(2n-1)}\wedge\alpha)\nonumber \\ \nonumber && +i_K(C^{(2n-1)}-R^{-2}i_{\alpha}C^{(2n-1)}\wedge \alpha)\wedge i_{\alpha}B\biggr]_{\mu_1\ldots \mu_{2n-1}}\\
i_{\beta}(i_{K^{+}}{\cal C}^{(2n)})_{\mu_1\ldots \mu_{n-2}}&\rightarrow&i_K(C^{(2n-1)}-R^{-2}i_{\alpha}C^{(2n-1)}\wedge\alpha)_{\mu_1\ldots \mu_{2n-2}}
\end{eqnarray}
We find that the D-brane charges obey the relations
\begin{eqnarray}
M_{(n)\mu_1\ldots \mu_n}&\leftrightarrow&-i_{\beta}N_{(n+1)\mu_1\ldots \mu_n}\\
i_{\alpha}M_{(n)\mu_1\ldots \mu_{n-1}}&\leftrightarrow&-N_{(n-1)\mu_1\ldots \mu_{n-1}}
\end{eqnarray}
Showing these relations is straightforward but long winded so the full details are not presented here. The important point is that the closed expressions satisfy qualitatively the same T-duality transformations as the branes do themselves, as we would hope.\\ 
\indent The NS-branes behave differently under T-duality from the D-branes. First we consider the F-strings: dualising an F-string transverse to its worldvolume will yield the F-string solution of the dual theory. We find that the charges obey the same relation, namely
\begin{eqnarray}
{\cal M}_{(1)\mu}\leftrightarrow-{\cal N}_{(1)\mu}
\end{eqnarray}
The NS5-branes on the other hand only T-dualise to one another if the isometry is along their worldvolumes. If the transformation is performed transverse to the worldvolume then one arrives at the KK monopole of the dual theory, with the T-duality isometry now coinciding with the Taub-NUT isometry. Once again these same relations occur between our charges. Specifically we find:
\begin{eqnarray}
i_{\alpha}{\cal M}_{(5)\mu_1\ldots \mu_4}&\leftrightarrow&-i_{\beta}{\cal N}_{(5)\mu_1\ldots \mu_4}\\
{\cal M}_{(5)\mu_1\ldots \mu_5}&\leftrightarrow&N_{(KK)\mu_1\ldots \mu_5}\\
M_{(KK)\mu_1\ldots \mu_5}&\leftrightarrow&-{\cal N}_{(5)\mu_1\ldots \mu_5}
\end{eqnarray}
Furthermore, T-dualising the KK monopole along its worldvolume leads to the KK-monopole of the dual theory. We have not showed that this occurs for our respective charges because we have not calculated how the fields $N^{(7)}$ and ${\cal N}^{(7)}$ transform when the T-duality is not performed along the isometry to which they are dual. This situation will be considered in \cite{Callister:2007}. However looking at the transformation rules of the bilinears we see that they at least transform as required and so the relation
\begin{eqnarray}
i_{\alpha}M_{(KK)\mu_1\ldots \mu_4}&\leftrightarrow&i_{\beta}N_{(KK)\mu_1\ldots \mu_4}
\end{eqnarray}
should hold. It is worth stressing that in this instance the Taub-NUT isometry is not the isometry along which the duality is occurring, hence the isometries appearing in the KK monopole charges do not transform according to the rules stated above.\\
\indent Finally the NS9-branes transform into one another by T-duality which must occur along the worldvolume since we are dealing with spacetime-filling branes here. We note that although we don't suppose (\ref{eq:IIA NS9-charge}) and (\ref{eq:IIB NS9-charge}) represent the full charges for the NS9-branes, they do satisfy the T-duality relation
\begin{eqnarray}
i_{\alpha}{\cal M}_{(9)\mu_1\ldots \mu_8}\leftrightarrow-i_{\beta}{\cal N}_{(9)\mu_1\ldots \mu_8}\end{eqnarray}
which at least supports the notion that the terms we have so far could well be correct.\\
\indent We thus see that all our expressions for the brane charges satisfy the same T-duality relations as the branes themselves. However, this alone is not enough to show that the actual charges themselves satisfy these T-dual relations. Recall that the functional form of our charges is gauge dependent, hence it is not enough to simply state the expression for the charge, one must also state the gauge conditions for the Lie derivatives that occur. When T-dualising the potentials using the rules stated above we are in effect using a specific mapping between a gauge choice in one theory to a gauge choice in the dual theory. Therefore, when mapping the charges to one another, we must also ensure that the gauge conditions map to one another.\\
\indent Recall that for IIB and the massless IIA theory, our gauge conditions required the Lie derivative of all the potentials to vanish with respect to the isometry generated by $K^{+}$ and $K$ respectively. A proper treatment of the T-duality relations of the charges should therefore check that all these conditions dualise into one another. Looking at the IIA condition ${\cal L}_KC^{(2n+1)}=0$ as an example, we find that this maps over to:
\begin{eqnarray}
  -{\cal L}_{K^{+}}(i_{\beta}{\cal C}^{(2n+2)})+{\cal L}_{K^{+}}({\cal C}^{(2n)}+{\cal R}^{-2}i_{\beta}{\cal C}^{(2n)}\wedge \beta)\wedge i_{\beta}{\cal B}=0
\end{eqnarray}  
which is consistent with our IIB conditions bearing in mind that ${\cal L}_{K^{+}}{\cal R}$ and ${\cal L}_{K^{+}}\beta$ both vanish. Such consistencies will be found between all the gauge conditions presented in the two theories though we do not present the details here. Hence the T-dual relations between our expressions for the charges actually do apply to the charges themselves in general, as we would hope. 

\subsection{Massive case}
In \cite{Bergshoeff:1996ui} the massive T-duality rules were determined which map the IIB theory to the massive IIA theory as opposed to the massless one. We do not fully consider these rules here but merely discuss the qualitative role they play with our charges. It is not consistent to include a mass parameter in the IIB theory so at first sight it is not clear what $m$ in the IIA theory should map to. It turns out that whereas with the massless T-duality rules where the potentials are chosen to be independent of the isometry direction, for a massive T-duality the mass parameter manifests itself in the IIB theory as a dependence of the potentials on the isometry direction. This is in accordance with the idea of generalised reductions discussed in \cite{Lavrinenko:1997qa,Meessen:1998qm}. To see why this is the case we simply have to observe that the general rule for T-dualising a field strength from IIA to IIB is given by:
\begin{eqnarray}
F^{(2n)}&\rightarrow&i_{\beta}{\cal F}^{(2n+1)}+{\cal F}^{(2n-1)}\wedge i_{\beta}{\cal B}\nonumber \\ &&+{\cal R}^{-2}i_{\beta}{\cal F}^{(2n-1)}\wedge i_{\beta}{\cal B}\wedge \beta
\end{eqnarray}
Therefore, Interpreting the IIA mass parameter $m$ as $F^{(0)}$ we find that it T-dualises to $i_{\beta}{\cal F}^{(1)}$. Ordinarily from the IIB perspective we would assume this quantity to be zero since it is equal to ${\cal L}_{\beta}{\cal C}^{(0)}$ but we see that generally it must be equal to $m$. This has the knock-on effect of giving all the higher rank potentials a dependence on the isometry proportional to $m$ as well. Note that since $m$ is a constant, ${\cal L}_{\beta}{\cal F}^{(1)}$ still vanishes as it must.\\
\indent Since the structure of the massless and massive IIA charges is the same, the only thing extra to consider when dualising the massive charges to the IIB theory is the massive gauge conditions for ${\cal L}_K$. These will amount to having a set of gauge conditions for ${\cal L}_{K^+}$ in the IIB theory which will be proportional to the IIB version of $m$, namely ${\cal L}_{\beta}{\cal C}^{(0)}$. We therefore conclude that the `massive' IIB charges simply involve a set of relations for ${\cal L}_{K^+}$ and ${\cal L}_{\beta}$ for which the IIB charges, including also the charges contracted with $\beta$, for example $i_{\beta}N_{(n)}$, are closed. We do not calculate these relations in this paper but it should be straight forward to calculate them from T-dualising the IIA relations. Note that equation (\ref{eq:IIB G8}) was only calculated for the case where ${\cal L}_{\beta}$ of the potentials vanished.  For the more general case extra terms may need to be added. 

\section{Conclusion}
In this paper we have considered the relation between the SUSY algebras and 1/2-BPS states for the maximal 10 and 11 dimensional supergravities. We constructed
expressions from bilinear forms, made from a Killing spinor, and gauge
potentials which were shown to be closed for curved supersymmetric backgrounds.
From these expressions the charge structure of the SUSY algebra in supersymmetric curved spacetimes can be deduced.\\
\indent These generalised charges displayed several features. Firstly
by construction, in flatspace they are simply related to the charges appearing in the flatspace SUSY algebra. Also, it was shown that the 11
dimensional and IIA charges are related in the same manner as the branes via
dimensional reduction, and similarly for the IIA and IIB charges by T-duality.
Finally, in calculating these charges it was found that the brane tensions automatically appear as part of the charges and the potentials to which the branes couple always appear in a similar fashion. This supports the association between these charges and the corresponding 1/2-BPS states.\\
\indent As well as considering the standard M-, NS- and D-branes, we also
considered the KK monopoles and M9-branes. In order to find both these charges
one had to make use of the isometries inherently present in their spacetime
solutions to construct an additional algebraic Killing spinor equation.
Considering the M9-brane and D8-brane forced us to consider the massive
supergravity theories. In this instance we found that the expressions for the
massless charges were still valid but the gauge conditions for the potentials had to
be generalised to the massive case.\\
\indent Further work is currently being carried out on the `exotic' branes that
appear in theses theories and will appear in the forthcoming paper \cite
{Callister:2007}.
 
\section*{Acknowledgements}
A.K. was funded by the Isle of Man Govt. DJS was supported in part by PPARC.

\appendix

\section{Conventions}\label{sec:conventions}
We denote the 11 dimensional objects by a hat, whilst the corresponding objects in the 10 dimensional theories are unhatted. All indices are unhatted. Orthonormal indices are denoted by Roman characters, whereas spacetime co-ordinate indices are denoted by characters from the Greek alphabet from the middle of the alphabet. Greek characters from the beginning of the alphabet are used to denote spinor indices. We do not distinguish between the 11 and 10 dimensional cases since we do not mix indices in this paper, thus an 11 dimensional object is assumed to have 11 dimensional indices etc. An isometry co-ordinate is labelled by a Roman letter at the end of the alphabet, usually $z$. For these co-ordinates an underline will be used to denote when we are working with an orthonormal basis.\\
\indent We use metrics with signature $(-,+,\ldots,+)$, and our antisymmetric symbol is defined (in a $D=d+1$ spacetime) by
\begin{eqnarray}
\epsilon_{01\ldots d}=+1
\end{eqnarray}
Our inner product convention is defined by
\begin{eqnarray}
(i_{\omega}F)_{\mu_1\ldots \mu_{p}}=\frac{1}{q!}\omega^{\nu_1\ldots \nu_q}F_{\nu_1\ldots \nu_q \mu_1\ldots \mu_{p}}
\end{eqnarray}
whilst our Hodge dual by
\begin{eqnarray}
(\ast F)_{\mu_1\ldots \mu_p}=\frac{\sqrt{|g|}}{p!}\epsilon_{\mu_1\ldots \mu_p}^{\phantom{\mu_1\ldots \mu_p}\nu_1\ldots \mu_{D-p}}F_{\nu_{1}\ldots \nu_{D-p}}
\end{eqnarray}
Combinations of $\Gamma$ matrices are assumed antisymmetrised, i.e.
\begin{eqnarray}
\Gamma_{\mu_1\ldots \mu_p}=\Gamma_{[\mu_1\ldots \mu_p]}
\end{eqnarray}
where there is a factor of $\frac{1}{p!}$ in our definition for anti-symmetrisation.\\
\indent When we work in an orthonormal basis adapted to an isometry such as $\alpha$, and parametrised by a co-ordinate such as $z$, the vielbeins are defined as 
\begin{eqnarray}
e^{\underline{z}}_\mu=R^{-1}\alpha_{\mu}& \qquad e^a_{z}=0 \qquad &e^{\underline{z}}_{z}=R\nonumber \\
e^{z}_{a}=-R^{-2}\alpha_{\mu}e^{\mu}_{a}& \qquad e^{\mu}_{\underline{z}}=0 \qquad &e^{z}_{\underline{z}}=R^{-1}
\end{eqnarray}
where $e^{\mu}_{a}$ and $e^{a}_{\mu}$ are still only defined implicitly, and $|\alpha|^2=R^2$.
\section{IIA from 11-dimensions} \label{sec:reducs}
Here we present the reduction rules of the 11-dimensional theory to IIA with our definitions and conventions. Firstly, in the appropriate co-ordinate system, the metric decomposes as  
\begin{eqnarray}
d\hat{s}^2_{(1,10)}&=&e^{-\frac{2}{3}\phi}ds^2_{(1,9)}+e^{\frac{4}{3}\phi}(dz+C^{(1)}
_{\mu}dx^{\mu})^2
\end{eqnarray}
where $z$ is the co-ordinate that parametrises the isometry we are reducing along.\\ 
\indent We have the following decomposition rules for the `standard' potentials
\begin{eqnarray}
\hat{A}&=&C^{(3)}+B\wedge dz\\
\hat{C}&=&B^{(6)}-(C^{(5)}-\frac{1}{2}C^{(3)}\wedge B)\wedge dz
\end{eqnarray}
For an isometry $\hat{\alpha}$ and its dual potential $i_{\hat{\alpha}}\hat{N}_{(8)}$, the decomposition rules differ depending on whether the reduction takes place along $\hat{\alpha}$ or not. In the case where the reduction is along $\hat{\alpha}$ then we have the following rules

\begin{eqnarray}
\hat{\alpha}&=&e^{\frac{4}{3}\phi}C^{(1)}+e^{\frac{4}{3}\phi}dz\\
i_{\hat{\alpha}}\hat{N}^{(8)}_{\mu_1\ldots \mu_7}&=&(-C^{(7)}+\frac{1}{3!}C\wedge (B)^2)_{\mu_1\ldots \mu_7}
\end{eqnarray}
whereas in the instance that the reduction occurs along a different isometry, the (partial) set of rules are given by
\begin{eqnarray}
\hat{\alpha}&=&e^{-\frac{2}{3}\phi}\alpha+e^{\frac{4}{3}\phi}i_{\alpha}C^{(1)}C^{(1)}+e^{\frac{4}{3}\phi}i_{\alpha}C^{(1)}dz\\
i_{\hat{\alpha}}\hat{N}^{(8)}_{\mu_1\ldots \mu_6z}&=&(i_{\alpha}N^{(7)}-\frac{1}{3}i_{\alpha}(C^{(3)}\wedge i_{\alpha}C^{(3)}\wedge B))_{\mu_1\ldots \mu_6}
\end{eqnarray}
Note that in this case this means that the radius of the isometry characterised by $\hat{\alpha}_z$ reduces as
\begin{eqnarray}
\hat{R}^2= e^{-\frac{2}{3}\phi}R^2+e^{\frac{4}{3}\phi}(i_{\alpha}C^{(1)})^2
\end{eqnarray}
The 10-form potential reduces as
\begin{eqnarray}
i_{\hat{\alpha}}\hat{A}^{(10)}_{\mu_1\ldots \mu_9}&=&(-C^{(9)}+\frac{1}{4!}C^{(3)}\wedge (B)^3)_{\mu_1\ldots \mu_9}
\end{eqnarray}
The spinors reduce according to 
\begin{eqnarray}
\hat{\epsilon}&=&e^{-\frac{1}{6}\phi}\epsilon
\end{eqnarray}
from which one can infer the reduction rules for the bilinear forms. These are most easily represented in the orthonormal frame defined above and are given by
\begin{eqnarray}
\hat{K}_{(1)}&=&\exp(-2\phi/3)K+\exp(-\phi/3)X\hat{e}^{\underline{z}}\\
\hat{\omega}_{(2)}&=&\exp(-\phi)\Omega+\exp(-2\phi/3)\tilde{K}\wedge\hat{e}^{\underline{z}}\\
\hat{\Sigma}_{(5)}&=&\exp(-2\phi)\Sigma+\exp(-5\phi/3)Z\wedge\hat{e}^{\underline{z}}\\
\hat{\Lambda}_{(6)}&=&\exp(-7\phi/3)\Lambda+\exp(-2\phi)\tilde{\Sigma}\wedge\hat{e}^{\underline{z}}\\
\hat{\Pi}_{(9)}&=&\exp(-10\phi/3)\Pi+\exp(-3\phi)\Psi\wedge\hat{e}^{\underline{z}}\\
\hat{\Upsilon}_{(10)}&=&\exp(-11\phi/3)\Upsilon+\exp(-10\phi/3)\tilde{\Pi}\wedge\hat{e}^{\underline{z}}
\end{eqnarray}

\section{Summary of charges}\label{sec:summary}

Here we collect all of the charges for easy reference.
\subsection{11-dimensional SUGRA}
For 11 dimensional SUGRA we find the charges for the M2-brane, M5-brane, KK monopole and M9-brane respectively to be given by
\begin{eqnarray}
\hat{L}_{(2)}&=&\hat{\omega}+i_{\hat{K}}\hat{A}\\
\hat{L}_{(5)}&=&\hat{\Sigma}+i_{\hat{K}}\hat{C}+\hat{A}\wedge(\hat{\omega}+\frac{1}{2}i_{\hat{K}}\hat{A})\\
\nonumber \hat{L}_{(KK)}&=&\hat{R}^2\hat{\Lambda}+i_{\hat{\alpha}}\hat{\Lambda} \wedge \hat{\alpha}-i_{\hat{K}}(i_{\hat{\alpha}}\hat{N}^{(8)}-\frac{1}{3!}\hat{A}\wedge (i_{\hat{\alpha}}\hat{A})^2)-i_{\hat{\alpha}}\hat{\omega}\wedge (i_{\hat{\alpha}}\hat{C}\\ &&+\frac{1}{2}\hat{A}\wedge i_{\hat{\alpha}}\hat{A})+i_{\hat{\alpha}}\hat{L}_{(5)}\wedge i_{\hat{\alpha}}\hat{A}-\frac{1}{2}\hat{L}_{(2)}\wedge (i_{\hat{\alpha}}\hat{A})^2\\
\nonumber i_{\hat{\alpha}}\hat{L}_{(9)}&=&\hat{R}^2i_{\hat{\alpha}}\hat{\Pi}+i_{\hat{K}}(i_{\hat{\alpha}}\hat{A}^{(10)}-\frac{1}{4!}\hat{A}\wedge (i_{\hat{\alpha}}\hat{A})^3)\\ \nonumber  &&-i_{\hat{\alpha}}\hat{\omega}\wedge (i_{\hat{\alpha}}\hat{N}^{(8)}-\frac{1}{3!}\hat{A}\wedge (i_{\hat{\alpha}}\hat{A})^2)+\hat{L}_{(KK)}\wedge i_{\hat{\alpha}}\hat{A}\\ &&-\frac{1}{2}i_{\hat{\alpha}}\hat{L}_{(5)}\wedge (i_{\hat{\alpha}}\hat{A})^2+\frac{1}{3!}\hat{L}_{(2)}\wedge (i_{\hat{\alpha}}\hat{A})^3
\end{eqnarray}
Where the following gauge conditions have been chosen:
\begin{eqnarray}
{\cal L}_{\hat{K}}\hat{A}&=&\hat{m}i_{\hat{\alpha}}\hat{A}\wedge i_{\hat{\alpha}}\hat{L}_{(2)}\\
{\cal L}_{\hat{K}}C&=&-\hat{m}i_{\hat{\alpha}}\hat{C}\wedge i_{\hat{\alpha}}\hat{L}_{(2)}+\hat{m}\hat{L}_{(KK)}\\
{\cal L}_{\hat{K}}(i_{\hat{\alpha}}\hat{N}^{(8)})&=&0\\
{\cal L}_{\hat{K}}(i_{\hat{\alpha}}\hat{A}^{(10)})&=&0
\end{eqnarray}

\subsection{Type IIA SUGRA}
For type IIA SUGRA we find the D-brane charges to be given by
\begin{eqnarray}
M_{(0)}&=&e^{-\phi}X-i_KC^{(1)}\\
M_{(2)}&=&e^{-\phi}\Omega+\tilde{K}\wedge C^{(1)}+i_KC^{(3)}+M_{(0)}B\\
M_{(4)}&=&e^{-\phi}Z-\tilde{K}\wedge
C^{(3)}-i_KC^{(5)}+M_{(2)}\wedge
B-\frac{1}{2}M_{(0)}(B)^2\\
\nonumber M_{(6)}&=&e^{-\phi}\Lambda+\tilde{K}\wedge
C^{(5)}+i_KC^{(7)}+M_{(4)}\wedge B-\frac{1}{2}M_{(2)}\wedge (B)^2\\ &&+\frac{1}{3!}M_{(0)}(B)^3\\
\nonumber M_{(8)}&=&e^{-\phi}\Psi-i_KC^{(9)}-\tilde{K}\wedge
C^{(7)}+M_{(6)}\wedge B-\frac{1}{2}M_{(4)}\wedge (B)^2\\
&&+\frac{1}{3!}M_{(2)}\wedge (B)^3-\frac{1}{4!}M_{(0)}(B)^4 
\end{eqnarray}
while for the  F-string and NS5-brane we have
\begin{eqnarray}
{\cal M}_{(1)}&=&\tilde{K}+i_KB\\
\nonumber {\cal M}_{(5)}&=&e^{-2\phi}\Sigma+e^{-\phi}(Z\wedge
C^{(1)}+\Omega\wedge C^{(3)}+XC^{(5)})+i_KB^{(6)}\\ &&+\tilde{K}\wedge C^{(1)}\wedge
C^{(3)}-i_KC^{(1)}C^{(5)}+\frac{1}{2}i_KC^{(3)}\wedge C^{(3)}
\end{eqnarray}
and finally the KK monopole charge is
\begin{eqnarray}
\nonumber M_{(KK)}&=&e^{-2\phi}R^2\tilde{\Sigma}-e^{-2\phi}i_{\alpha}\tilde{\Sigma}\wedge\alpha+e^{-\phi}i_{\alpha}C^{(1)}i_{\alpha}\Lambda\\ \nonumber &&-i_K(i_{\alpha}N^{(7)})+\frac{1}{2}i_K(B\wedge (i_{\alpha}C^{(3)})^2)\\ \nonumber &&+i_{\alpha}(e^{-\phi}\Omega+\tilde{K}\wedge C^{(1)})\wedge(i_{\alpha}C^{(5)}-i_{\alpha}C^{(3)}\wedge B)\\ \nonumber &&+i_{\alpha}\tilde{K}(i_{\alpha}B^{(6)}+\frac{1}{2}C^{(3)}\wedge i_{\alpha}C^{(3)})+i_{\alpha}{\cal M}_{(5)}\wedge i_{\alpha}B\\ &&+i_{\alpha}M_{(4)}\wedge i_{\alpha}C^{(3)}-M_{(2)}\wedge i_{\alpha}C^{(3)}\wedge i_{\alpha}B\nonumber \\ &&-\frac{1}{2}{\cal M}_{(1)}\wedge(i_{\alpha}C^{(3)})^2
\end{eqnarray}
In this instance the gauge conditions are understood to be 
\begin{eqnarray}
{\cal L}_KB&=&0\\
{\cal L}_KC^{(2n-1)}&=&-\frac{1}{(n-1)!}m{\cal M}_{(1)}\wedge (B)^{(n-1)}\\
{\cal L}_KB^{(6)}&=&-m{\cal M}_{(1)} \wedge C^{(5)}+\frac{1}{2}m{\cal M}_{(1)} \wedge C^{(3)}\wedge B\nonumber \\ && +mM_{(6)}\\
{\cal L}_K(i_{\alpha}N^{(7)})&=&-mi_{\alpha}{\cal M}_{(1)}\wedge i_{\alpha}C^{(7)}+mi_{\alpha}B\wedge i_{\alpha}M_{(6)}\nonumber \\ &&-m\frac{1}{2}i_{\alpha}({\cal M}_{(1)} \wedge (B)^2)\wedge i_{\alpha}C^{(3)}
\end{eqnarray}

\subsection{Type IIB SUGRA}
For type IIB SUGRA we find the D-brane charges to be given by
\begin{eqnarray}
N_{(1)}&=&e^{-\varphi}K^{12}-{\cal C}^{(0)}K^{-}+i_{K^{+}}{\cal C}^{(2)}\\
N_{(3)}&=&e^{-\varphi}\Phi^{12}+K^{-}\wedge
{\cal C}^{(2)}-i_{K^{+}}{\cal C}^{(4)}+N_{(1)}\wedge B\\
\nonumber N_{(5)}&=&e^{-\varphi}\Sigma^{12}-K^{-}\wedge
{\cal C}^{(4)}+i_{K^{+}}{\cal C}^{(6)}+N_{(3)}\wedge B\\
&&-\frac{1}{2}N_{(1)}\wedge (B)^2\\
\nonumber N_{(7)}&=&e^{-\varphi}\Pi^{12}+K^{-}\wedge
{\cal C}^{(6)}-i_{K^{+}}{\cal C}^{(8)}+N_{(5)}\wedge B\\
&&-\frac{1}{2}N_{(3)}\wedge (B)^2+\frac{1}{3!}N_{(1)}\wedge
(B)^3\\
\nonumber N_{(9)}&=&e^{-\varphi}\Omega^{12}-K^{-}\wedge
{\cal C}^{(8)}+i_{K^{+}}{\cal C}^{(10)}+N_{(7)}\wedge B\\\nonumber
&&-\frac{1}{2}N_{(5)}\wedge (B)^2+\frac{1}{3!}N_{(3)}\wedge
(B)^3\\ &&-\frac{1}{4!}N_{(1)}\wedge (B)^4
\end{eqnarray}
while the F-string and NS5-brane charges are given by
\begin{eqnarray}
{\cal N}_{(1)}&=&K^{-}-i_{K^{+}}B\\
\nonumber {\cal N}_{(5)}&=&e^{-2\varphi}\Sigma^{-}+e^{-\varphi}({\cal C}^{(0)}\Sigma^{12}+\Phi^{12}\wedge {\cal C}^{(2)}+K^{12}\wedge {\cal C}^{(4)})\\ \nonumber &&-i_{K^+}{\cal B}^{(6)}+\frac{1}{2}K^{-}\wedge
{\cal C}^{(2)}\wedge {\cal C}^{(2)}-{\cal C}^{(0)}K^{-}\wedge {\cal C}^{(4)}\\ &&+i_{K^{+}}{\cal C}^{(2)}\wedge
{\cal C}^{(4)}
\end{eqnarray}
and for the KK monopole we have
\begin{eqnarray}
\nonumber N_{(KK)}&=&e^{-2\varphi}{\cal R}^2\Sigma^{+}-e^{-2\varphi}i_{\beta}\Sigma^{+}\wedge \beta-i_K(i_{\beta}{\cal N}^{(7)}) +e^{-\varphi}i_{\beta}\Sigma^{12}\wedge i_{\beta}{\cal C}^{(2)}\\ \nonumber &&+e^{-\varphi}i_{\beta}\Phi^{12}\wedge i_{\beta}{\cal C}^{(4)}+e^{-\varphi}i_{\beta}K^{12}\wedge i_{\beta}{\cal C}^{(6)}-i_{\beta}{\cal N}_{(5)}\wedge i_{\beta}{\cal B}\\ \nonumber &&+i_{\beta}K^-(-i_{\beta}{\cal B}^{(6)}+{\cal C}^{(2)}\wedge i_{\beta}{\cal C}^{(4)}-{\cal C}^{(0)}i_{\beta}{\cal C}^{(6)})\\ \nonumber &&+K^-\wedge i_{\beta}{\cal C}^{(4)}\wedge i_{\beta}{\cal C}^{(2)}+ \frac{1}{2}i_K(i_{\beta}{\cal C}^{(4)})\wedge i_{\beta}{\cal C}^{(4)}\nonumber \\ &&-i_K(i_{\beta}{\cal C}^{(2)})\wedge i_{\beta}{\cal C}^{(6)}
\end{eqnarray}
In this paper we only consider the IIB theory that T-dualises to the massless IIA theory in which case we have ${\cal L}_{K^+}$ of all the potentials vanishing. Furthermore when an isometry is present in the case of the KK monopole or for the purposes of T-dualising to IIA, we also choose the Lie derivatives along the isometry to vanish for all the potentials. More general conditions will exist that will make the theory T-dual to the massive IIA theory, but we did not consider these in this paper.

\end{document}